\documentclass{jfm}

\usepackage{graphicx}
\usepackage{natbib}
\usepackage{amssymb}
\usepackage{amsmath}
\usepackage{latexsym}

\usepackage{bm}
\usepackage[squaren]{SIunits}
\usepackage{lipsum}
\usepackage{color}
\usepackage{mathrsfs}
\usepackage[hypertexnames=false,colorlinks=true,linkcolor=blue,citecolor=blue]{hyperref}

\graphicspath{{eps/}{pdf/}}

\ifCUPmtlplainloaded \else
  \checkfont{eurm10}
  \iffontfound
    \IfFileExists{upmath.sty}
      {\typeout{^^JFound AMS Euler Roman fonts on the system,
                   using the 'upmath' package.^^J}%
       \usepackage{upmath}}
      {\typeout{^^JFound AMS Euler Roman fonts on the system, but you
                   dont seem to have the}%
       \typeout{'upmath' package installed. JFM.cls can take advantage
                 of these fonts,^^Jif you use 'upmath' package.^^J}%
      }
  \else
  \fi
\fi

\ifCUPmtlplainloaded \else
  \checkfont{msam10}
  \iffontfound
    \IfFileExists{amssymb.sty}
      {\typeout{^^JFound AMS Symbol fonts on the system, using the
                'amssymb' package.^^J}%
       \usepackage{amssymb}%
         \let\leq=\leqslant
         \let\geq=\geqslant
      }{}
  \fi
\fi

\ifCUPmtlplainloaded \else
  \IfFileExists{amsbsy.sty}
    {\typeout{^^JFound the 'amsbsy' package on the system, using it.^^J}%
     \usepackage{amsbsy}}
    {}
\fi

\newsavebox{\astrutbox}
\sbox{\astrutbox}{\rule[-5pt]{0pt}{20pt}}

\newcommand{\miller}[1]{\langle#1\rangle}

\title[Homoclinic snaking near the surface instability of a polarizable fluid]{Homoclinic snaking near the surface instability of a polarizable fluid}

\author[David J.B. Lloyd, Christian Gollwitzer, Ingo Rehberg, and Reinhard Richter]
{David J.B. Lloyd$^1$%
  \thanks{Email address for correspondence: d.j.lloyd@surrey.ac.uk},\ns
Christian Gollwitzer$^2$,  Ingo Rehberg$^2$,\break  and Reinhard Richter$^2$}

\affiliation{$^1$Department of Mathematics, University of Surrey, Guildford, GU2 7XH, UK\\[\affilskip]
$^2$Experimentalphysik V, Universit\"at Bayreuth, D-95440 Bayreuth, Germany}

\pubyear{2015}
\volume{xxx}
\pagerange{xxx--xxx}

\date{?; revised ?; accepted ?. - To be entered by editorial office}

\begin{document}

\maketitle







\begin{abstract}
We report on localized patches of cellular hexagons observed on the surface of a magnetic fluid in a vertical magnetic field. These patches are spontaneously generated by jumping into the neighborhood of the unstable branch of the domain covering hexagons of the Rosensweig instability upon which the patches equilibrate and stabilise. They are found to co-exist in intervals of the applied magnetic field strength parameter around this branch. We formulate a general energy functional for the system and a corresponding Hamiltonian that provides a pattern selection principle allowing us to compute Maxwell points (where the energy of a single hexagon cell lies in the same Hamiltonian level set as the flat state) for general magnetic permeabilities. Using numerical continuation techniques we investigate the existence of localized hexagons in the Young-Laplace equation coupled to the Maxwell equations. We find cellular hexagons possess a Maxwell point providing an energetic explanation for the multitude of measured hexagon patches. Furthermore, it is found that planar hexagon fronts and hexagon patches undergo homoclinic snaking corroborating the experimentally detected intervals. Besides making a contribution to the specific area of ferrofluids, our work paves the ground for a deeper understanding of homoclinic snaking of 2D localized patches of cellular patterns in many physical systems.
\end{abstract}

\maketitle

\section{Introduction}
Spatial localization is proving to be key to understanding various coherent features seen in fluids comprising Taylor-Couette flow~\citep{abshagen2010}, plane Couette flow~\citep{schneider2010,chantry2014}, binary fluid convection~\citep{moses1987,batiste2006,mercader2013}, and transition-to-turbulence~\citep{pringle2014}. See also \citet{knobloch2008,knobloch2015}, \citet{dawes2010} and \citet{descalzi2011} for broader reviews. In the Homoclinic Snaking scenario \citep{pomeau1986,woods1999,coullet2000} each alternating turn of a ``snake" in control parameter-phase space is correlated with the emergence of a further interior cell of the localized pattern. This scenario has proven to be an important mechanism for localization in more than hundred theoretical studies. However, experimental evidence is limited, so far, to buckling \citep{thompson2015}, a semiconductor laser  \citep{barbay2008hss}, vertically vibrated media~\citep{umbanhowar1996}, nonlinear optics~\citep{akhmediev2005}, gas discharges \citep{purwins2010}, a light valve \citep{haudin2011} and in the context of fluids, the Taylor-Couette flow \citep{schneider2010,abshagen2010}. A pending problem is a comparison of experiment and theory of homoclinic snaking for fully localized 2D patterns (not just structures that are localized in a single spatial direction) in a physical fluid system.

\begin{figure}
\centering
\includegraphics[width=\linewidth]{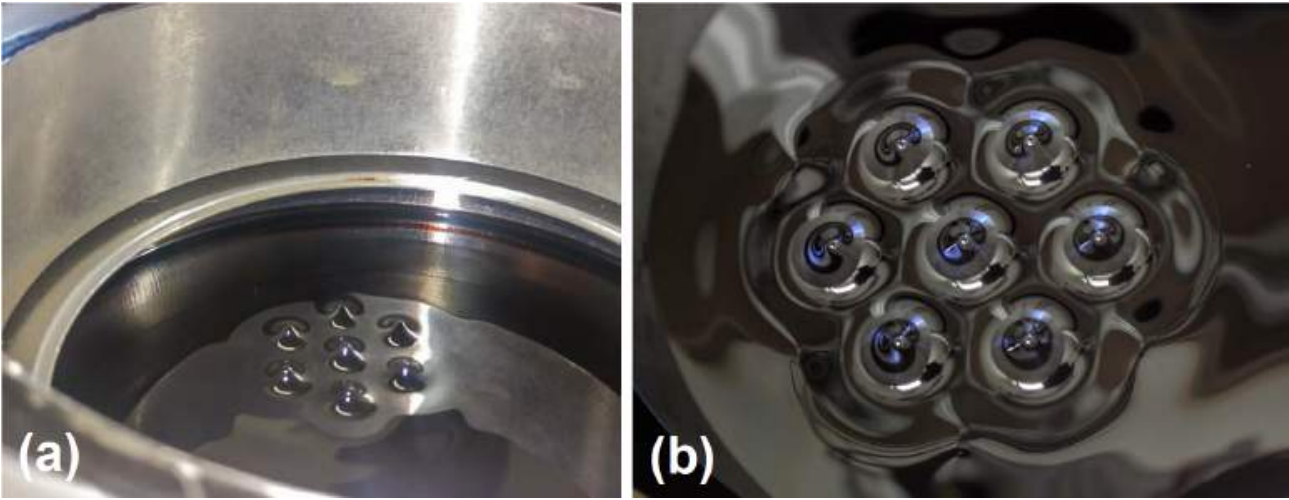}
\caption{A patch of seven ferrofluidic spikes seen together with the upper coil (a), and zoom in high density range (b). Photographs courtesy of Robin Maretzki (a) and Achim Beetz (b).
\label{f:7patch}}
\end{figure}

A step towards filling this gap is investigating the surface instability of an electrically or magnetically polarizable fluid, because spatial inhomogeneities and noise are very low. When a static layer of magnetic fluid is subjected to a vertical magnetic field, it becomes unstable when a critical value $B_c$ of the magnetic induction is surpassed. Due to this Rosensweig instability~\citep{cowley1967} a regular hexagonal pattern of cellular spikes emerges. By applying local magnetic pulses, localized radially symmetric spikes could be generated on the free-surface~\citep{richter2005}.
In this paper we demonstrate that various \emph{localized hexagon patches} like in figure\,\ref{f:7patch} emerge \emph{spon\-tan\-eously} if the \emph{homogeneous} field is switched into a regime where we expect to find homoclinic snaking.
So far the modeling of localized spikes is limited to finite element simulations \citep{lavrova2008,cao2014}
and a  simple reaction-diffusion model \citep{richter2005}. However, the Rosensweig instability involves a free-surface and so the justification of reaction-diffusion theory is problematic.
Solving the full dynamic equations, we present numerical evidence that the localized hexagon patches undergo homoclinic snaking similar to that observed in reaction-diffusion theory~\citep{woods1999,coullet2000}.

The article is outlined as follows. In \S\ref{s:exp_setup}, we describe the experimental setup,
 which is followed by a characterization of the ferrofluid (\S\ref{s:ferro_properties}) and the experimental results (\S\ref{s:exp_results}). We then introduce in \S\ref{s:gov.eqs} the mathematical model for the experiment that we numerically solve and present in \S\ref{s:numerical_results} the numerical results. Section~\ref{s:comparison} then compares experimental and theoretical results. We finish with conclusions and outlook (\S\ref{s:conclusion}).

\section{Experimental setup}\label{s:exp_setup}
\begin{figure}
\centering
\includegraphics[width=0.8\linewidth]{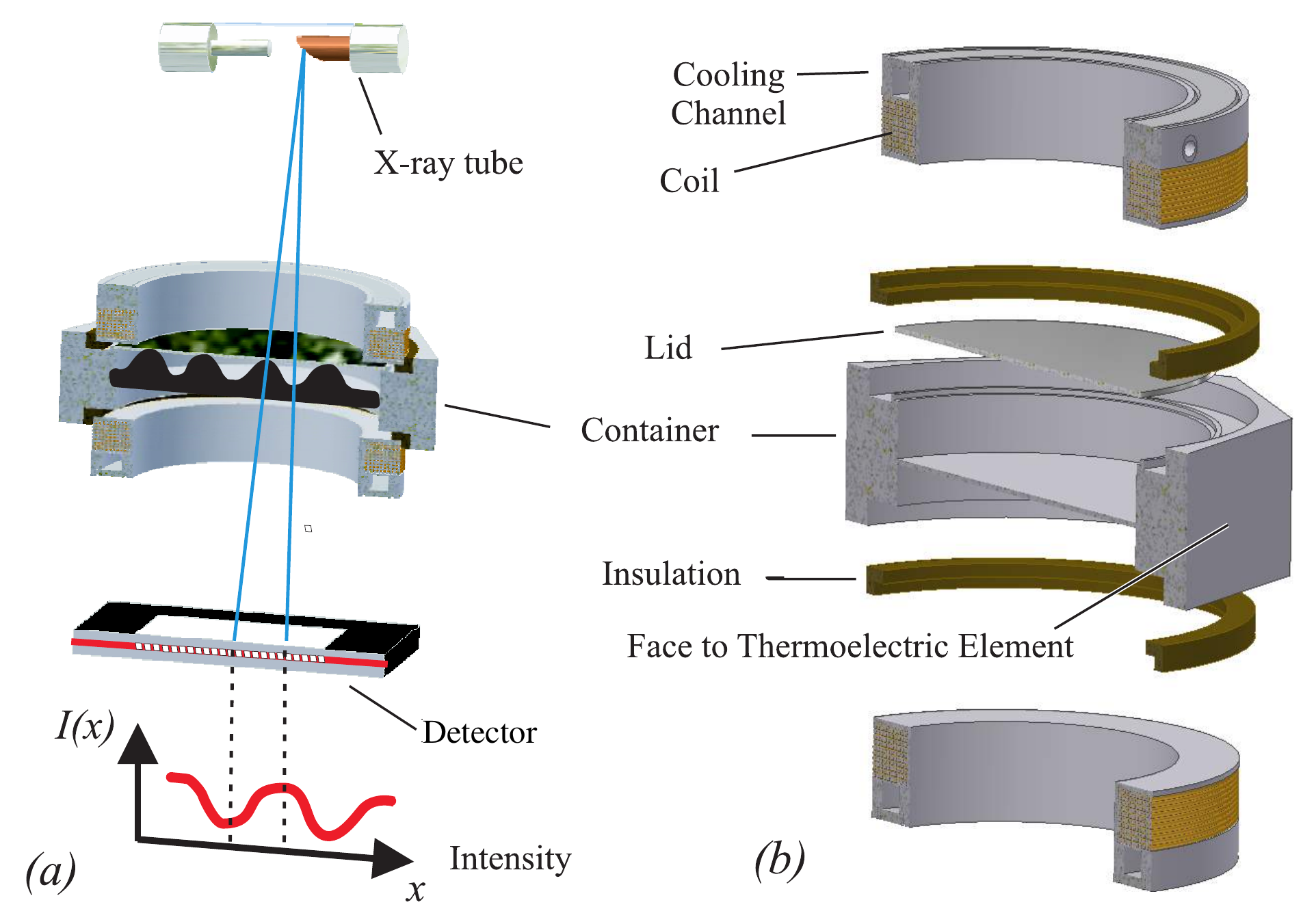}
\caption{Setup of the apparatus for measuring the Rosensweig
instability. Scheme of the assembled setup (\textit{a}), and
exploded view of the container with the coils (\textit{b}).}
\label{f:setup}
\end{figure}

Our experimental setup is sketched in figure\,\ref{f:setup}(a).
An X-ray point source emits radiation vertically
from above through the container with the ferrofluid, which is placed midway
between a Helmholtz pair of coils. Underneath the container,
an X-ray camera records the radiation passing through the layer of ferrofluid.
The intensity at each pixel of the detector is directly related to the height of
the fluid above that pixel. Therefore, the full surface topography can be
reconstructed after calibration \citep{richter2001,gollwitzer2007}.

The container, which holds the ferrofluid sample, is depicted in
figure\,\ref{f:setup}(\textit{b}). It is a regular octagon machined from
aluminium with a side length of $77\,\mathrm{mm}$ and two concentric inner
bores with a diameter of $140\,\mathrm{mm}$. These circular holes are carved from
above and below, leaving only a thin base in the middle of the vessel with a
thickness of $2\,\mathrm{mm}$. On top of the octagon, a circular lid of aluminium is placed,
which closes the hole from above (see figure~\ref{f:setup}\textit{b}).
Each side of the octagon is equipped with a
thermoelectric element QC-127-1.4-8.5MS from Quick-Ohm.
These are powered by a $1.2\,\mathrm{kW}$ Kepco KLP-20-120  power supply. The hot side
of the thermoelectric elements is connected to water cooled heat exchangers. The
temperature is measured at the bottom of the aluminium container with a
Pt100 resistor. The temperature difference between the center and the edge of the bottom plate does not exceed $0.1\,$K at temperature $10.0\,^\circ$C measured at the edge of the vessel.
A closed loop control, realized using a computer and programmable interface devices, holds the temperature $\theta$ of the vessel constant with an accuracy of $10\,\mathrm{mK}$.

The container is surrounded by a Helmholtz pair of coils, thermally isolated
from the vessel with a ring made from a flame resistant composite material
(FR-2). As shown in figure~\ref{f:setup} the diameter of the coils is intentionally smaller than the diameter of the vessel in order to introduce a magnetic ramp. With this arrangement the magnetic field strength falls off towards the border of the vessel, where it reaches $80\,\%$ of its value in the center.
The ramp is introduced in order to minimize the differences between
the amplitudes obtained in a finite and an infinite geometry. For details of
the magnetic ramp see \citet{knieling2010}. Filling the container to a height of $5\,\mathrm{mm}$ with ferrofluid enhances the magnetic induction in comparison with the empty coils
for the same current $I$. Therefore $B(I)$ is measured immediately beneath the
bottom of the container, at the central position, and serves as the control
parameter in the following.

\section{Properties of the Magnetic Fluid}\label{s:ferro_properties}
For the experiments we use the ferrofluid APG\,E32 from Ferrotec Co. Table \ref{tab:apge32.parameters} lists the important material parameters. The density $\rho$
was measured utilizing an electronic density meter (DMA 4100) from Anton Paar.
The surface tension was measured by means of a ring tensiometer (Lauda TE 1) and  corroborated by the pendant drop method (Dataphysics OCA 20).

\begin{table}
\begin{center}
\begin{tabular}{lllll}
Quantity                                        &                & \textnormal{Value}    & Error       & Unit                 \\ \hline
Density			                                & $\rho$         & 1168.0 & $\pm 1$     & $\mathrm{kg\,m^{-3}}$ \\
Surface tension                                 & $\sigma$       & 30.9   & $\pm 5$     & $\mathrm{mN\,m^{-1}}$ \\
Viscosity at $10\;^\circ\mathrm{C}$             & $\eta$         & 4.48   & $\pm 0.1$   & $\mathrm{Pa\,s}$\\

Initial permeability from (\ref{e:vis})         & $\mu_\mathrm{r,i}$   & 4.57  & $\pm 0.005$ & \\

Saturation magnetization from (\ref{e:vis})     & $M_\mathrm{S}$ & 23.15 & $\pm 0.8$   & $\mathrm{kA\,m^{-1}}$ \\

Critical induction from \,(\ref{e:M.c})         & $B_\mathrm{c,\infty}$  & 10.83  & $\pm 0.1$   & $\mathrm{mT}$\\[1mm]

Critical induction from experiment              & $B_\mathrm{c,exp}$   & 11.23  & $\pm 0.1$   & $\mathrm{mT}$\\

\end{tabular}
\caption{Properties of the magnetic fluid  APG\,E32 (Lot~G090707A) from Ferrotec Co.}
\label{tab:apge32.parameters}
\end{center}
\end{table}

The viscosity $\eta$ has been measured in a temperature range of $-5\;^\circ\mathrm{C}\leq \theta \leq 20\;^\circ\mathrm{C}$, using a commercial rheometer (MCR\,301, Anton Paar) with a shear cell featuring a cone-plate geometry. At room temperature, the magnetic fluid with a viscosity of $\eta=2\,\mathrm{Pa\cdot s}$ is $2000$ times more viscous than water. The value of $\eta$ can be increased by a factor of $9$ when the liquid is cooled to $-5\;^\circ\mathrm{C}$. The temperature dependent viscosity data can be well described by the Vogel-Fulcher law
\citep{rault2000}
\begin{equation}
\eta=\eta_0\exp\left(\frac{\psi}{\theta-\theta_0}\right),
\label{eq:vogelfulcher}
\end{equation}
with $\eta_0=0.48\,\mathrm{mPa\cdot s}, \psi=1074\,\mathrm{K},$ and
$\theta_0=-107.5\;^\circ\mathrm{C}$, as reported in detail by \cite{gollwitzer2010}.
For the present measurements, we chose a temperature of
$\theta=10\;^\circ\mathrm{C}$, where the viscosity amounts to
$\eta=4.48\,\mathrm{Pa\cdot s}$ according to equation (\ref{eq:vogelfulcher}).

\begin{figure}
\centering
\includegraphics[width=\linewidth]{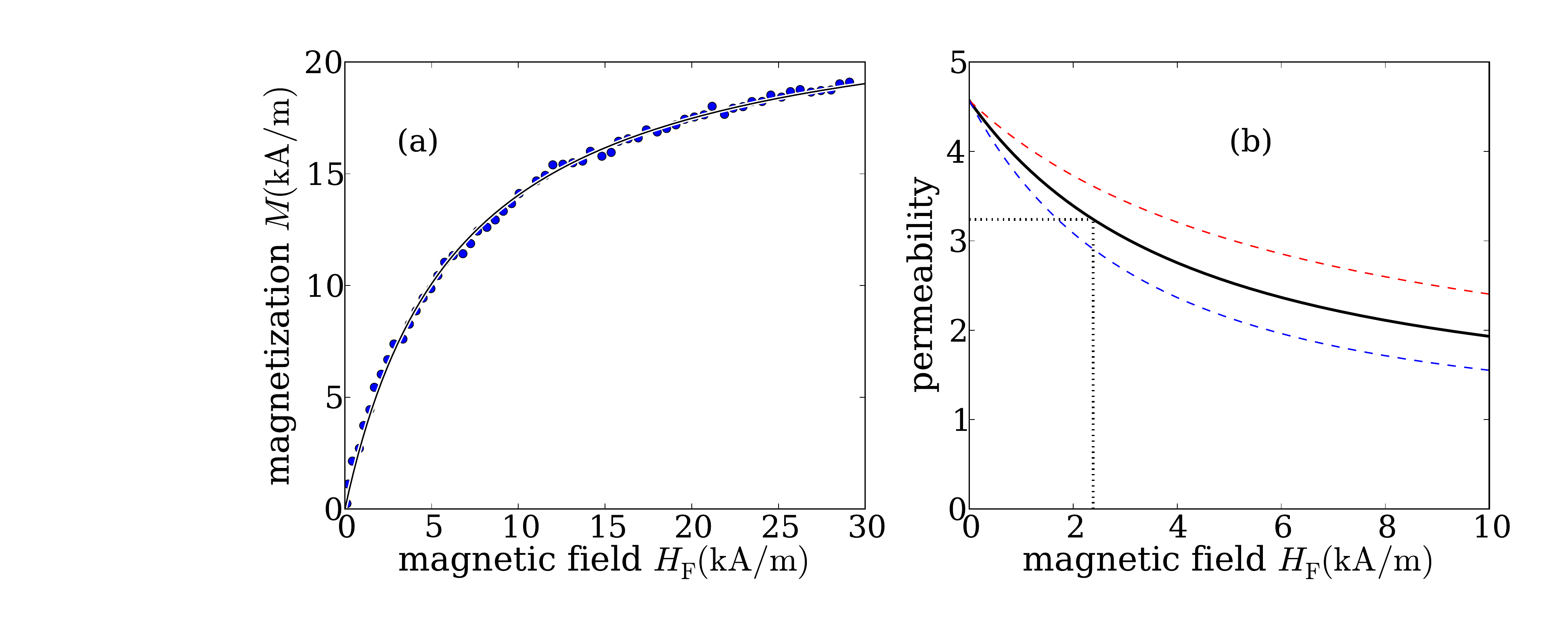}
\caption{(\textit{a}) Measured magnetisation curve versus the field in the fluid, and (\textit{b}) related chord- (red dashed), tangent- (blue dashed) and effective (black solid line) permeabilities. The dotted line marks $\mu_\mathrm{eff}$ at $H_c$.  }
\label{f:magnetisation}
\end{figure}

The magnetization, $\mathbf{M}(\mathbf{H}_F)$, where $\mathbf{H}_F$ is the magnetic field in the fluid, has been measured by means of a fluxmetric magnetometer (Lake\-shore Model 480).
A spherical cavity with a volume of $1\,\mathrm{ml}$ was constructed from perspex$\circledR$ and filled up to the brim with the magnetic fluid. By means of x-rays we checked for a bubble free filling. Figure \ref{f:magnetisation}(a) presents the measured data together with a fit (solid line) by equation
\begin{equation}
M(H_\mathrm{F}) = M_\mathrm{S} \frac{H_\mathrm{F}}{H_\mathrm{F} + \frac{M_\mathrm{S}}{\mu_\mathrm{r,i}-1}},
\label{e:vis}
\end{equation}
utilized by \cite{Froehlich1881} and \cite{vislovich1990}, where $M = |\mathbf{M}|$, $M_\mathrm{S}$ is the saturation magnetisation, $H_\mathrm{F}$ is the magnitude of the applied magnetic field in the fluid and $\mu_{\mathrm{r,i}}$ is the initial magnetic permeability.

Fitting the data with (\ref{e:vis}) we obtain for the initial magnetic permeability $\mu_\mathrm{r,i}= 4.57$, and for the saturation magnetisation $M_\mathrm{S}=23.15\,\mathrm{kA/m}$. Equation (\ref{e:vis}) is used to calculate the chord permeability
\begin{equation}
\mu_\mathrm{ch} \equiv M/H_\mathrm{F}+1
\end{equation}
and the tangent permeability
\begin{equation}
\mu_\mathrm{ta} \equiv \partial M/\partial H_\mathrm{F}+1 = \frac{(\mu_\mathrm{r,i}-1)}{\left(1+ (\mu_\mathrm{r,i}-1) \frac{H_\mathrm{F}}{M_\mathrm{S}}\right)^2} + 1,
\end{equation}
which are displayed in figure \ref{f:magnetisation}(b) by a dashed red (blue) line, respectively. The effective permeability
\begin{equation}
\mu_\mathrm{eff}=\sqrt{\mu_\mathrm{ta}(H_\mathrm{F}) \cdot \mu_\mathrm{ch}(H_\mathrm{F})}
\label{e.eff}
\end{equation}
is the geometric mean of both permeabilities \citep{cowley1967} and marked in figure \ref{f:magnetisation}(b) by the black solid line.

By means of $\mu_\mathrm{eff}(M)$, the material parameters $\sigma$, $\rho$ and the gravitational acceleration $g=9.81 \mathrm{m/s^2}$ the critical magnetization for a semi-infinite layer of ferrofluid  is approximated by the implicit equation \citep{rosensweig1985}
\begin{equation}
M_c(M) = \frac{2}{\mu_0}\sqrt{1 + \frac{1}{\mu_\mathrm{eff}(M)} \sqrt{\rho g \sigma}}.
\label{e:M.c}
\end{equation}
By numerically solving the implicit equation we obtain $M_\mathrm{c}=6.2303\,\mathrm{kA/m}$, $H_\mathrm{c}=2.3872\,\mathrm{kA/m}$, and $B_\mathrm{c}=\mu_0 (H_\mathrm{c} + M_\mathrm{c})= 10.83 \,\mathrm{mT}$. Note that this value is just an approximation for a semi-infinite layer of ferrofluid. Revisiting figure \ref{f:magnetisation}(b), we see that $\mu_\mathrm{eff}$ drops considerably for increasing field. At the critical field we find $\mu_\mathrm{eff}(H_c)=3.2$ (cf. the dotted line). The critical induction found by fitting amplitude equations to the experimental data \citep{gollwitzer2010} is about 4\% higher (cf. Table \ref{tab:apge32.parameters}).

We stress that the mapping between a linear and a nonlinear $M(H)$ by (\ref{e.eff}) is only valid in the linear regime of pattern formation, i.e. for the onset of the instability, but not for the fully developed nonlinear patterns.

\section{Experimental results}\label{s:exp_results}
\begin{figure}
\centering
\includegraphics[width=1\linewidth]{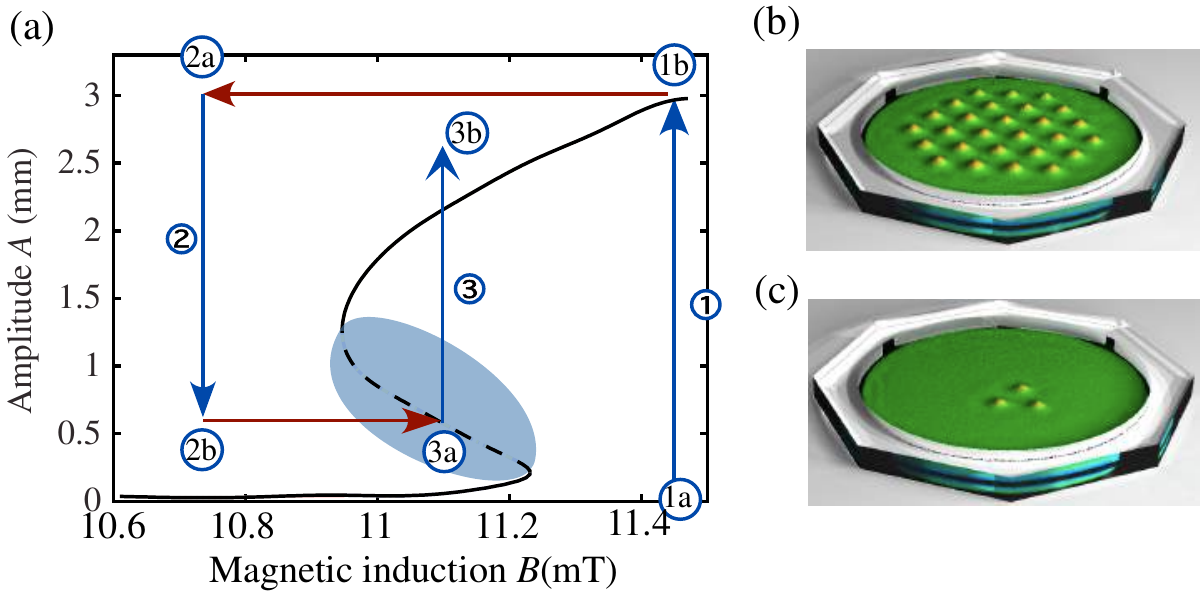}
\caption{(a) The bifurcation diagram (black line) of this pattern is obtained by fitting equation (11) in \citet{gollwitzer2010} to about 170\,000 measured data. Via the path indicated by the arrows one obtains localized patches, as shown in (c). $B$ is kept fixed during (1) for a time delay of $\tau_1$, then $B$ is switched to (2) for a time delay of $\tau_2$ before finally being switched to (3) until equilibrium is reached at point (3b). Panel (b)presents a reconstruction of the height field of the domain covering hexagons at point (1a). Panel (c) gives a reconstruction of the height field of a three-spike patch at point (3b) in the bifurcation diagram.\label{f:exp_protocol}}
\end{figure}

In the investigated regime the plain surface and domain covering regular up-hexagonal (i.e. hexagons whose maximum amplitude is postive) patterns are bistable  ($10.95\,\mathrm{mT}\leq B\leq 11.23\,\mathrm{mT}$), as shown by the bifurcation diagram in figure~\ref{f:exp_protocol}\,(a), which stems from a fit to about 170\,000 measured data points \citep{gollwitzer2010}. Panel (b) displays an experimental realization of the domain covering hexagonal pattern observed at the upper branch, the infinite extension of which has the symmetry of the dihedral group $\mathbb{D}_6$.

Due to the high viscosity of the magnetic fluid the formation of Rosensweig patterns takes 60 seconds. This allows us to vary the magnetic induction $B$ (the control parameter) and the pattern amplitude $A$ almost independently. In this way we can enter the region around the unstable branch of the instability, where homoclinic snaking is suspected, from the side, as sketched in figure~\ref{f:exp_protocol}.

Starting from a flat layer, $B$ is always switched to an over critical value $B_1=11.45\,\mathrm{mT}$, cf.\, position (1a) in figure~\ref{f:exp_protocol}. With time the amplitude of the domain covering hexagon pattern increases along path (1). After $\tau_1=60.0\,\mathrm{s}$ this pattern is stabilized (1b). Then $B$ is switched to a value $B_2=10.74\,\mathrm{mT}$ before the fold of the hexagons and the magnetic fluid is allowed to relax from (2a) towards the flat state. Before the magnetic fluid flattens completely, $B$ is switched at (2b) to a value $B_3$ in the neighborhood {(3a) of the unstable branch of the imperfect transcritical bifurcation \citep[cf. equations\,(1,2) in][]{gollwitzer2010}}, marked by a dashed line and shaded region in figure~\ref{f:exp_protocol}. Now the pattern relaxes on path\,(3a) to (3b) where the height of the patch finally settles down (equilibrates) to a value approximately $1.5\,\mathrm{mm}$ higher (depending on the size of the patch i.e., for larger patches the height tends to that for the domain-covering hexagons) than the domain covering hexagons. A movie \protect\citep{video3patch} shows the birth of a three spike patch.
Each patch was at least stable for 120 seconds, as recorded by the measuring program.
Moreover, we have checked the long term stability of exemplary patches after generation and observed no decay for more than two hours.

\begin{figure}
\centering
\includegraphics[width=0.8\linewidth]{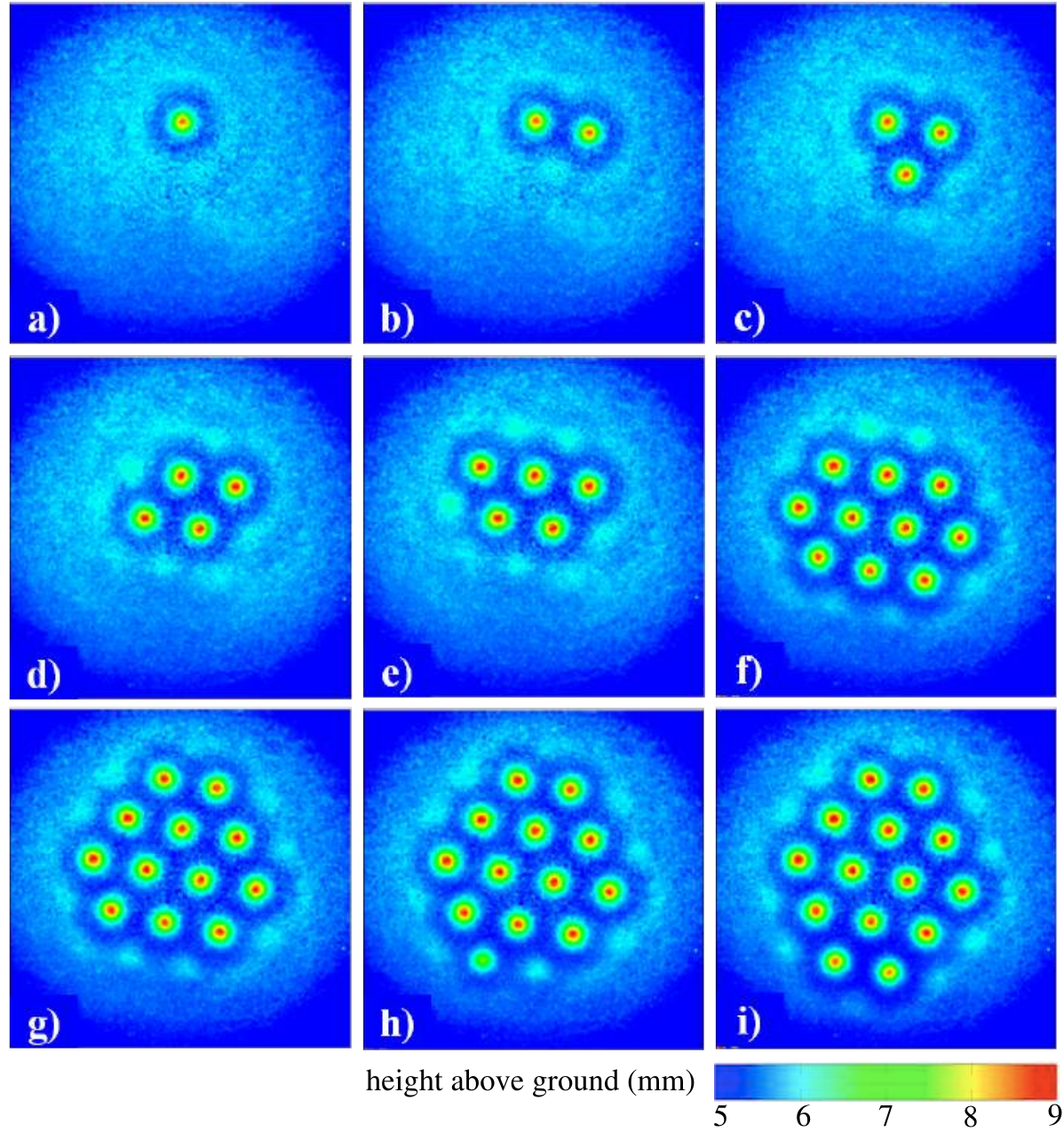}
\caption{Localized states for $\tau_2=10.0\,\mathrm{s}$ at (a)\,B=11.0725\,mT, (b)\,\,B=11.0857\,mT, (c)\,\,B=11.1120\,mT, (d)\,\,B=11.1648\,mT, (e)\,\,B=11.1780\,mT, (f)\,\,B=11.1911\,mT, (g)\,\,B=11.2043\,mT, (h)\,\,B=11.2175\,mT, and (i)\,\,B=11.2307\,mT.
\label{f:exp_loc}}
\end{figure}

Starting at the edge of the bistable region defined by the fold of the domain-covering cellular hexagon spikes and increasing $B_3$, we obtain a sequence of localized states ranging from solitary spikes to patches containing multiple spikes/hexagon cells on an otherwise flat fluid surface with two to 24 cells, as shown in figure~\ref{f:exp_loc} for {$\tau_2=10.0\,\mathrm{s}$}.
The height of the spikes/cells of a patch are not uniform; we observe that spikes near the centre of the patch are higher than those on the edge of the patch.
The distance between spikes within a patch is approximately the same as the spatial period of the domain-covering hexagon pattern. Remarkably, the spikes do not spread out in space as time increases despite their magnetic dipole-dipole repulsion, originating from the fact that the spikes are magnetized in parallel. In this way the patches are not an ensemble of weakly interacting solitary spikes in the sense of \,\citet{richter2005}.

Moreover it is interesting to note the mechanism by which hexagon cells are added to large patches as $B_3$ is increased. Going from panel~\ref{f:exp_loc}(f) to (i), we see that spikes are added in the middle of the lower front in order to form a completed patch with dihedral symmetry $\mathbb{D}_2$. It is also possible to see completed patches with $\mathbb{D}_6$-symmetry, as shown in Fig.\,\ref{f:7patch}.

Similar patches have been uncovered in the 2D Swift-Hohenberg-Equation as a signature of homoclinic snaking \citep[cf.][]{lloyd2008,escaff2009,kozyreff2013}. Also the mechanism governing the growth of a patch, as observed above, resemble the ones described by these authors.

{As shown in figure~\ref{f:exp_step}, the number of spikes in a patch increases mono\-tonical\-ly with $B_3$, with clearly defined plateaus.
The data related with figure\,\ref{f:exp_loc} are marked by $\bigtriangledown$ and have been recorded with a delay of $\tau_2 = 10.0\,\mathrm{s}$. Reducing $\tau_2$ to 9.5\,s ($\Box$),
$8.5\,\mathrm{s}$\,($\bigcirc$), and $7.5\mathrm{s}$\,($\bigtriangleup$) shifts the curves to lower magnetic inductions: Because the starting amplitudes are higher, a lower $B_3$ is sufficient to generate a localised patch.
Most importantly the prominent plateaus at 3, 10, 12, 14,\dots cells exist for all these values of $\tau_2$.
For the very short delay of $\tau_2=1.0\,\mathrm{s}$, the curve ($\diamond$) hardly exhibits plateaus (apart from the plateau at 27 spikes, which is the maximal size of a hexagonal pattern in the setup). That is because for $\tau_2=1.0\,\mathrm{s}$, the regular hexagonal pattern (see (1b) in figure \ref{f:exp_protocol}) has hardly decayed, and we can not sensitively explore the phase space where homoclinic snaking occurs.

\begin{figure}
\centering
\includegraphics[width=0.8\linewidth]{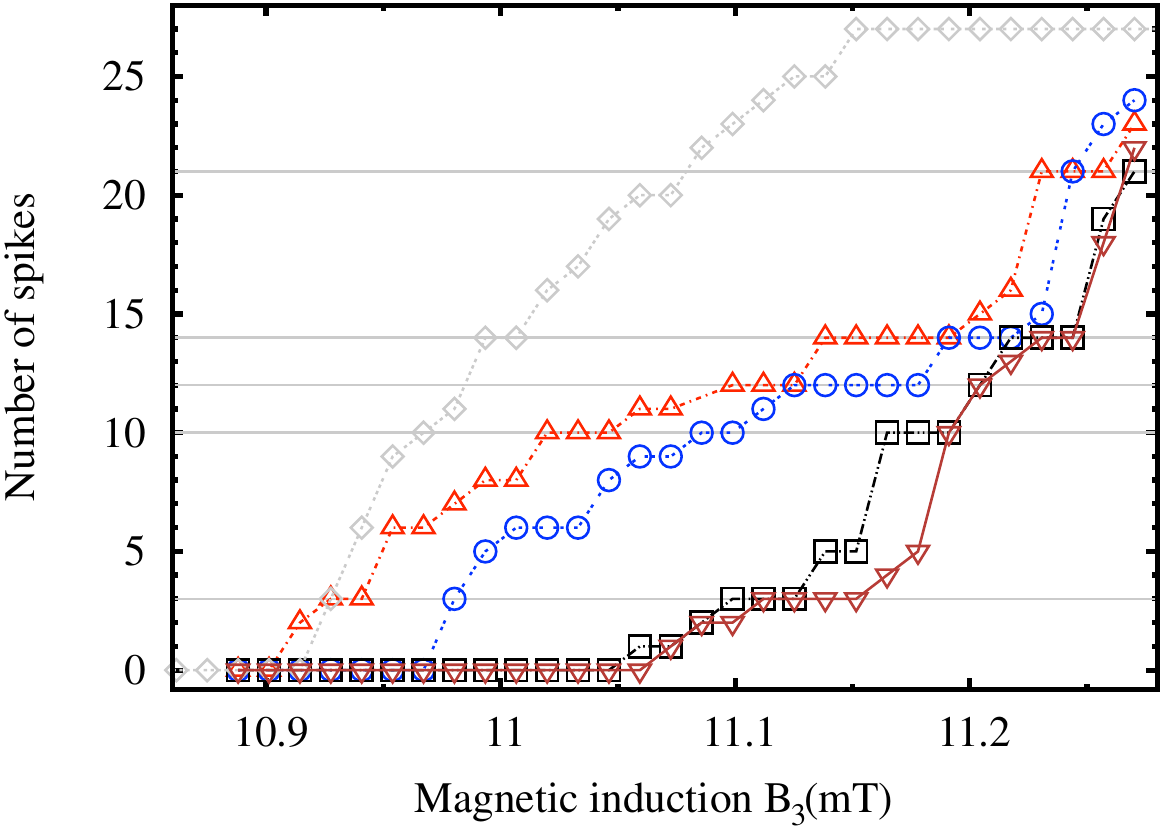}
\caption{The number of spikes vs $B_3$ for different delays $\tau_2 = 1.0\,\mathrm{s}$\, ($\diamond$), $7.5\mathrm{s}$\,($\bigtriangleup$), $8.5\,\mathrm{s}$\,($\bigcirc$), $9.5\,\mathrm{s}$\,($\Box$), and $10.0 \,\mathrm{s}$\,($\bigtriangledown$) exhibits plateaus. Plateaus at 3, 10, 12, 14 and 21 spikes,
where patches are forming complete patches, are marked by horizontal lines.
\label{f:exp_step} }
\end{figure}

\section{Governing equations} \label{s:gov.eqs}
In order to elucidate the snaking process we numerically solve the full 3D dynamical equations.
The field in the fluid is modeled by Maxwell's equations with no free currents and a Young-Laplace equation for the free-surface. Since there are no free currents, the magnetic field in the fluid, $\mathbf{H}_F$, can be written in terms of a gradient field $\mathbf{H}_F=h\mathbf{\hat{z}}-\nabla\phi$, where $h$ is the magnitude of the applied uniform magnetic field strength in the fluid in the vertical direction $\mathbf{\hat{z}}$ and the non-uniform part of the field is described by $-\nabla\phi$. The magnetisation
is given by the linear equation $M(H_F)=(\mu_r -1)H_\mathrm{F}$, where
$H_\mathrm{F} = |\mathbf{H_\mathrm{F}}|$ and $\mu_r=$const is the magnetic permeability of the magnetic fluid relative to that of free-space, $\mu_0$. Thus the magnetic induction follows from the linear relation $\mathbf{B}=\mu_r\mu_0\mathbf{H}_F$.

Of course a linear magnetisation law can only be a rough approximation of the real $M(H_\mathrm{F})$ measured in the experiment (\S\ref{s:ferro_properties}). Despite that, previous analytical studies which are based on a linear $M(H_\mathrm{F})$, can successfully predict all regular planforms on the ferrofluids and their underlying bifurcations  \citep[see][]{gailitis1977,friedrichs2001,friedrichs2002,bohlius2011aer}.
Likewise in our numerical study we aim to catch the decisive qualitative features seen in the experiment, and to elucidate whether they are connected to homoclinic snaking.

The magnetic field in the air above the magnetic fluid is given by $\mathbf{H}_A=\mu_rh\mathbf{\hat{z}}-\nabla\chi$ and the steady state free surface is given by $z=\zeta(x,y)$ with $\zeta=0$ corresponding to the flat state. The equations are nondimensionalized using the critical wavenumber
$k_c= \sqrt{\rho g/\sigma}$ for the length scale and $\sqrt{\rho g\sigma}$ for the unit of pressure, where $\rho$ is the fluid density and $\sigma$ is the surface tension. Furthermore, the amplitude $h$ of the uniform field strength inside the magnetic fluid is rescaled using $\mu_r$
\begin{equation}\label{e:curledH}
\mathscr{H} = \sqrt{\frac{\mu_0 \mu_r}{2(\mu_r+1)}}(\mu_r-1)h,
\end{equation}
such that the instability occurs at $\mathscr{H_\mathrm{c}}=1$. We note, when comparing equation (\ref{e:curledH}) with the related (5.5) of the work by \cite{silber1988}, we have rescaled with respect to $\mu_0$. The bifurcation parameter is then
\[
\epsilon = \mathscr{H}-1.
\]
We note that $\epsilon$ depends on $\mu_r$. For a static magnetic fluid in a vertical field, the dimensionless equations are
\begin{subequations}\label{e:sys}
\begin{align}\label{e:sys_a}
\nabla\left(p^* + z - \frac12(\mu_r-1)|\mathbf{H}_F|^2\right) &\mathbf{=0},\; z\in(-D,\zeta),\\ \label{e:sys_b}
\nabla^2\phi &= 0,\; \mbox{$z\in(-D,\zeta)$},\\
\quad \nabla^2\chi &= 0,\; \mbox{$z\in(\zeta,D)$}, \label{e:sys_c}
\end{align}
\end{subequations}
where $D$ is the depth of the fluid and $p^*$ is an effective pressure in the magnetic fluid. The first of these equations~(\ref{e:sys_a}) is the stationary Navier-Stokes equation for the magnetic fluid, and the other equations~(\ref{e:sys_b},\ref{e:sys_c}) are derived from Maxwell's equations assuming a linear magnetisation law; see~ \citet{cowley1967}. We consider a domain above the fluid to a height $D$ for computational ease. At the interface $z=\zeta$ the boundary conditions are given by
\begin{subequations}\label{e:comp}
\begin{align}\label{e:comp_a}
p^* + \frac12(\mu_r-1)^2\left(\mathbf{H}_F\cdot\mathbf{\hat{n}} \right)^2 = 2\kappa,\\
\left[\mathbf{B}\cdot\mathbf{\hat n} \right]_-^+ = 0,\qquad\left[\mathbf{H}\times\mathbf{\hat n}\right]_-^+ =\mathbf{0},
\end{align}
\end{subequations}
where $[\;\cdot\;]_-^+$ is the value of the jump across the interface of the quantity in the brackets,
$\mathbf{\hat{n}}$ is the unit normal to the interface, and $\kappa$ is the mean curvature of the surface:
\[
\mathbf{\hat n} = \frac{\mathbf{\hat z}-\nabla\zeta}{(1+|\nabla\zeta|^2)^{1/2}},\qquad \kappa=\frac{\nabla \cdot\mathbf{\hat n}}{2},
\]
see~\citet{silber1988}. No-flux boundary conditions are imposed on the top and bottom of the domain i.e., $\phi_z(\mathbf{x},-D)=\chi_z(\mathbf{x},+D)=0$ where $\mathbf{x}=(x,y)$. Solving~(\ref{e:sys_a}) for the pressure, within a constant, $C$, at the free-surface yields
\begin{equation}
p^*=-\zeta+\frac12(\mu_r-1)|\mathbf{H}_F|^2+C,\qquad\mbox{at $z=\zeta$}.
\end{equation}
The pressure can then be eliminated and the boundary conditions written in terms of the fields $\zeta(x,y),\phi(x,y,z)$, and $\chi(x,y,z)$, as well as two parameters, $\mu_r$ and $\epsilon$.

The system of equations possesses a free-energy given by
\begin{align}
&\mathcal{E}(\zeta,\phi,\chi)=\int_{\Omega}\left[\frac12\left(\int_{\zeta(\mathbf{x})}^D|\mathbf{H}_{A}|^2\mathrm{d}z + \mu_r\int_{-D}^{\zeta(\mathbf{x})}|\mathbf{H}_{F}|^2\mathrm{d}z \right)-\frac12\zeta^2 \right.\nonumber\\
\label{e:energy}
&\left. + C\zeta+ \mu_r h(\chi_{z=D}-\phi_{z=-D}) - \sqrt{1+|\nabla\zeta|^2}-1 \right]\mathrm{d}\mathbf{x},
\end{align}
made up of magnetic, hydrostatic, surface energies, respectively,
with conservation of mass, boundary conditions at the top and bottom of the domain, and with the boundary condition at the interface $\phi(\mathbf{x},\zeta)-h\zeta(1-\mu_r) =\chi(\mathbf{x},\zeta)$ as constraints, where $\Omega\subset\mathbb{R}^2$ is the space domain of $\zeta(\mathbf{x})$ and $\mathbf{x}=(x,y)$. We note that \,\citet{gailitis1977} and \citet{bohlius2006} use other energy formulations without the boundary conditions and conservation of mass.  The constant $C$ is a Lagrange multiplier for the mass constraint: $\int_{\Omega}\zeta\mathrm{d}\mathbf{x}$ equal to a constant. Taking variations of $\mathcal{E}$ with respect to the free-surface $\zeta$ yields the interface  equation~(\ref{e:comp_a}), and taking variations with respect to $\phi$ and $\chi$, using the constraint, yields the Laplace equations~(\ref{e:sys_b},\ref{e:sys_c}) and the no-flux boundary conditions $\phi_z(\mathbf{x},-D)=\chi_z(\mathbf{x},+D)=0$. The advantage of this free-energy is that it represents the whole system and we can now numerically compute energies taking into account the magnetic fields.

In order to fix the pressure constant, $C$, we assume that the base state $(\chi,\phi,\zeta)=(0,0,0)$ solves the equations corresponding to the mass constraint $\int_{\Omega}\zeta\mathrm{d}\mathbf{x}=0$. We choose to study the large depth of magnetic fluid case and we set the depth, $D$, to be $10$. For $\mu_r=2$, this is yields a depth of 16.42 mm i.e., about 3 times the depth of the fluid in the experiment, which is sufficiently large that the domain covering hexagons do not change as the depth is increased. We have ignored the effect of the shallow layer in our numerics since one would have to solve for an additional magnetic field below the magnetic fluid and apply suitable compatibility conditions.

Carrying out a coordinate transformation to flatten out the free-surface~\citep{twombly1983} given by
\begin{equation}\label{e:coord_1}
\left(\begin{array}{c}x \\ y \\ z \end{array} \right) \rightarrow \left(\begin{array}{c}\hat x \\ \hat y \\ \hat z \end{array} \right) = \left(\begin{array}{c}x \\ y \\ \frac{z-\zeta}{D-\zeta}D \end{array} \right),\qquad \mbox{for $\zeta<z<D$},
\end{equation}
and
\begin{equation}\label{e:coord_2}
\left(\begin{array}{c}x \\ y \\ z \end{array} \right) \rightarrow \left(\begin{array}{c}\hat x \\ \hat y \\ \hat z \end{array} \right) = \left(\begin{array}{c}x \\ y \\ \frac{z-\zeta}{D+\zeta}D \end{array} \right),\qquad \mbox{for $-D<z<\zeta$},
\end{equation}
we can perform a Legendre transformation in order to define a Hamiltonian. Arbitrarily taking the $x$-direction, we define the momenta coordinates $\alpha = \partial \mathcal{E}/\partial\phi_x,\beta = \partial\mathcal{E}/\partial \chi_x $ and $\eta =\partial\mathcal{E}/\partial\zeta_x$. Then the Hamiltonian for the system is given by
\begin{align}\label{e:ham}
\mathcal{H}(\phi,\chi,\zeta,\alpha,\beta,\eta) = \int_{0}^D\alpha\phi_x\mathrm{d}z + \int_0^D\beta\chi_x\mathrm{d}z + \eta\zeta_x - \mathcal{E},
\end{align}
on the constraint manifold $\mathcal{M}=\{\phi(\mathbf{x},0)-h\zeta(1-\mu_r) - \chi(\mathbf{x},0)=0, \phi_x(\mathbf{x},0)-h\zeta_x(1-\mu_r) - \chi_x(\mathbf{x},0)=0\}$; see~\cite{groves2015} and \citet{groves1997} for the similar case of water waves.

The Hamiltonian~(\ref{e:ham}) is a constant along solutions in the $x$-direction and provides a wavelength selection principle \citep[see][for the same principle applied to the Swift-Hohenberg equation]{lloyd2008}. In particular, if we evaluate $\mathcal{H}$ along any localized planar (hexagon) front connecting to the trivial state $\zeta=0$ and the domain-covering hexagon pattern, we find that $\mathcal{H}=\mathcal{H}(\mathbf{0})$ along the limiting domain-covering hexagon pattern so that it belongs to the same level set. Generically, $\mathcal{H}$ will only equal $\mathcal{H}(\mathbf{0})$ at finitely many spatially periodic orbits in the family of spatially periodic patterns and will therefore serve as a selection principle that involves only the spatially periodic patterns; see~\cite{beck2009} for more details.

\section{Numerical results}\label{s:numerical_results}
The 3D equations~(\ref{e:sys_b},\ref{e:sys_c}) and~(\ref{e:comp_a}) are solved by first flatting out the free-surface using~(\ref{e:coord_1})-(\ref{e:coord_2})  so that the computational domains are fixed cuboids. We apply Neumann boundary conditions on the four sizes of $\Omega$ in order to remove the translational invariance in $x$ and $y$ that would cause problems with Newton Solvers. This choice of boundary conditions on $\Omega$ limits us to cellular hexagons, symmetric planar fronts and $\mathbb{D}_2$ symmetric patches involving cellular hexagons computed on the positive quadrant; see~\citet{lloyd2008} for more details.  To remove the continuous symmetry due to the magnetic potentials being defined only up to a constant, we add an extra parameter $\lambda$ to the Laplace equations (\ref{e:sys_b},\ref{e:sys_c}) to be solved along with $\phi,\chi$ and $\zeta$ and we append an extra equation \citep[see][]{twombly1983}
\begin{equation}\label{e:phase_con}
\chi(0,D) + \mu \phi(0,-D) = 0.
\end{equation}
We now have a well-posed boundary value problem with all the continuous symmetries removed  allowing us to use Newton methods.
We discretise in \textsc{matlab} using fourth-order central finite-differences (for planar fronts) or a pseudo-spectral Fourier method (for hexagon spikes and patches) in $x,y$-directions and Chebyshev pseudo-spectral method in $z$ with a Newton-GMRES method~\citep{kelley2003} and secant pseudo-arclength continuation~\citep{krauskopf2007}. For the Newton-GMRES scheme, one needs a pre-conditioner. For this we pre-compute a sparse incomplete LU-decomposition (with the Crout algorithm~\citep{saad1996}) of the linear system with a flat interface at onset and this is fixed for the continuation. The incomplete LU-decomposition is computed with a drop tolerance of 1$\times$10$^{-6}$ and we set a maximum of 20 iterations for GMRES with 40 Arnoldi restarts and a tolerance of $10^{-4}$. Typical spatial step sizes in $x$- and $y$-directions are 0.25-0.5 for the Fourier-psuedo spectral method and 20 Chebyshev collocation points in $z$ and a fixed arc length step size of $0.01$. The numerics are validated by comparing with normal form results by~\citet{silber1988}. In particular we check that the change of sub/supercriticality of rolls and squares occurs at the theoretically predicted values of $\mu_r;$ see the appendix for more details. Fold loci are found by carrying out parameter slices and then doing a linear interpolation about the fold values. The Maxwell point is computed by appending the Hamiltonian and Energy conditions to the system.

In the following we present the results obtained with the procedures sketched above. We start with the hexagon Maxwell curve (\S\ref{ss:hexagon.maxwell.curve}), which is succeeded by hexagon planar fronts (\S\ref{ss:hexagon.planar.fronts}), hexagon patches (\S\ref{ss:hexagon.patches}) and rhomboid patches (\S\ref{ss:rhomboid.patches})

\begin{figure}
\centering
\includegraphics[width=0.5\linewidth]{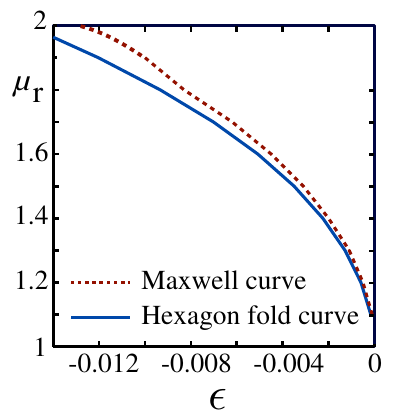}
\caption{Two parameter ($\epsilon,\mu_r$)-bifurcation diagram showing the existence of the primary fold of the domain covering hexagons and their corresponding Maxwell point. \label{f:maxwell}}
\end{figure}

\subsection{Hexagon Maxwell curve}\label{ss:hexagon.maxwell.curve}
Intuitively, one expects localized patches to exist when the energy of a single hexagon cell of the infinitely extended pattern has the same energy as the trivial flat state. This occurs at the Maxwell point: a hexagon with a wavelength selected by $\mathcal{H}$ has the same energy $\mathcal{E}(\mathbf{hex})=\mathcal{E}(\mathbf{0})$, where $\mathbf{hex}$ is a single hexagon cell. For \mbox{$\mu_r$} close to $1$, the hexagon Maxwell point was previously computed analytically by\,\citet{friedrichs2001} with an energy minimization principle for the wavelength selection. In order to find a Maxwell curve relevant for localized hexagon patterns one needs to use the Hamiltonian selection principle. We find that the Hamiltonian wavelength selection criterion only slightly changes the wavelength from that at onset.
In figure~\ref{f:maxwell}, we trace out the hexagon Maxwell curve varying both $\epsilon$ and $\mu_r$. We find that the Maxwell curve closely follows the fold of the domain covering hexagons, in particular for $\mu_r=2$ the Maxwell point is found to be $\epsilon\approx-0.013$. For $\mu_r$ close to $1$, the height of the hexagons becomes very small making it difficult for our numerics to capture since we only compute solutions to a relative error of 10$^{-4}$ and round-off error starts to dominate. We do believe though that the Maxwell curve should continue down to $\mu_r=1$.

\begin{figure}
\centering
\includegraphics{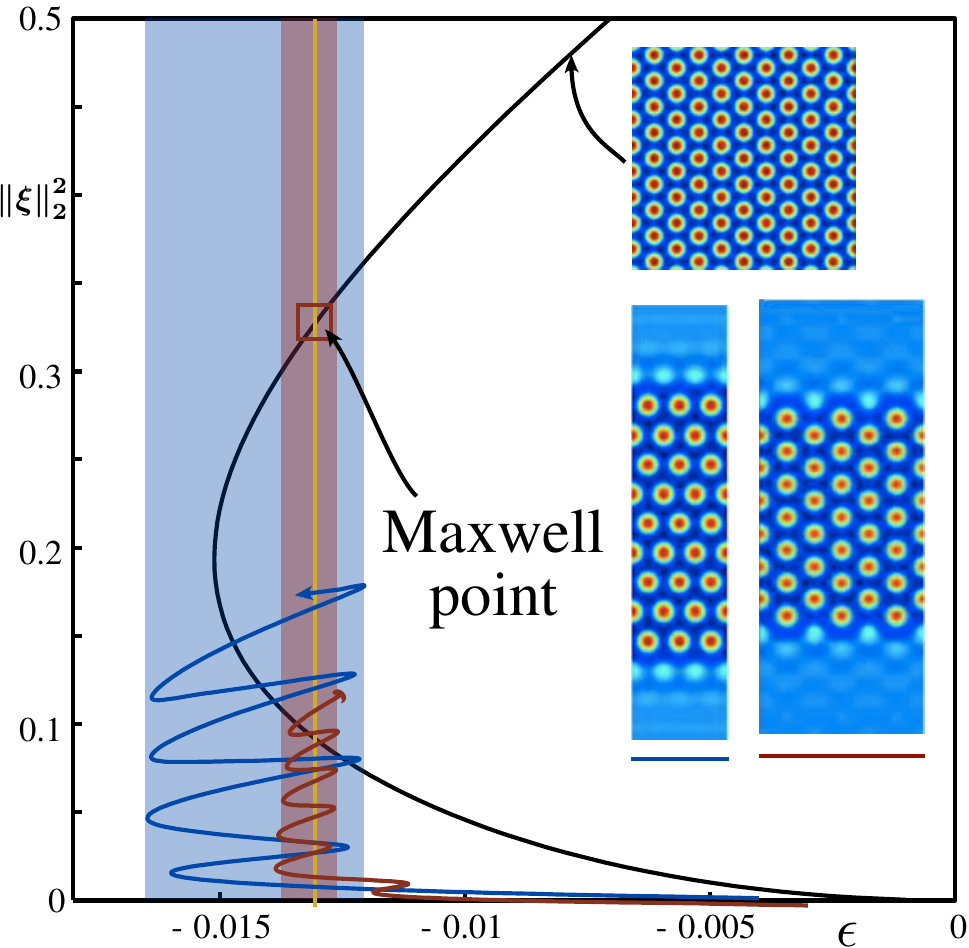}
\caption{L2 norm of the displacement of the interface of localized planar hexagon fronts undergo homoclinic snaking about the Maxwell point ($\Box$) for $\mu_r=2$. The $\langle10\rangle$-front (blue) is  snaking in a far larger region of parameter space than the $\langle11\rangle$-front (red). Here $\|\zeta\|_2^2 = \frac{1}{|\Omega|}\int_{\Omega}\zeta^2\mathrm{d}\mathbf{x}$ is the normalized L2 norm of the interface. \label{f:planar_fronts}}
\end{figure}
At the hexagon Maxwell curve, we expect to see stationary fronts between the flat state and domain covering hexagons since we are connecting states with the same energy. Infinitely-many different types of planar fronts are possible depending on the front's interface with respect to the underlying hexagonal lattice. These different fronts can be classified by their Bravais-Miller index which we denote by $\miller{n_1n_2}$ where $n_j$ are the reciprocals of the intercepts of the line with the lines $\mathbb{R}l_i$ generated by the lattice vectors $l_j$, assigning the reciprocal $n_j = 0$ if the line does not intersect $\mathbb{R}l_i$  \citep[see][]{ashcroft1976,lloyd2008}. It is these different types of fronts that form the sides of large hexagon patches.

\subsection{Hexagon planar fronts}\label{ss:hexagon.planar.fronts}
The works of~\citet{lloyd2008,escaff2009,kozyreff2013,dean2014}, suggest that one needs only to consider two main hexagon planar fronts, namely the $\miller{10}$ and $\miller{11}$ fronts, in order to understand snaking of fully localised patches. We note that in figure\,\ref{f:exp_loc}(f)-(i), we see that the patches are mostly made up of $\miller{10}$ fronts, with figure\,\ref{f:exp_loc}(f) also showing the left- and right-hand sides that are $\miller{11}$ fronts. Hence, we will concentrate on first understanding the $\miller{10}$ and $\miller{11}$ hexagon planar fronts.

Generically, we expect these hexagon fronts to remain pinned in an open region of parameter space where they undergo homoclinic snaking about the Maxwell point \citep[see][]{pomeau1986,beck2009}. This is confirmed by our numerical results presented in figure\,\ref{f:planar_fronts}: For $\mu_r=2$ indeed the $\miller{10}$- and $\miller{11}$-fronts snake about the Maxwell point at $\epsilon\approx-0.013$. We find, similar to that reported in other systems~\citep{lloyd2008,escaff2009,kozyreff2013,lloyd2013}, that the $\miller{10}$- front snakes over a larger region of parameter space than the $\miller{11}$-front. This gives a partial explanation as to why mostly patches with $\miller{10}$ sides are observed in our experiments as these live in a larger region of parameter space.

If the computational domain is commensurate with the domain-covering hexagons then we expect the snaking branches to terminate near the fold of the domain-covering hexagons; see~\cite{dawes2008}. In figure\,\ref{f:planar_fronts}, we have stopped the snaking branch early and indicate that the snaking should continue indefinitely for fronts on the plane.
\begin{figure}
\centering
\includegraphics{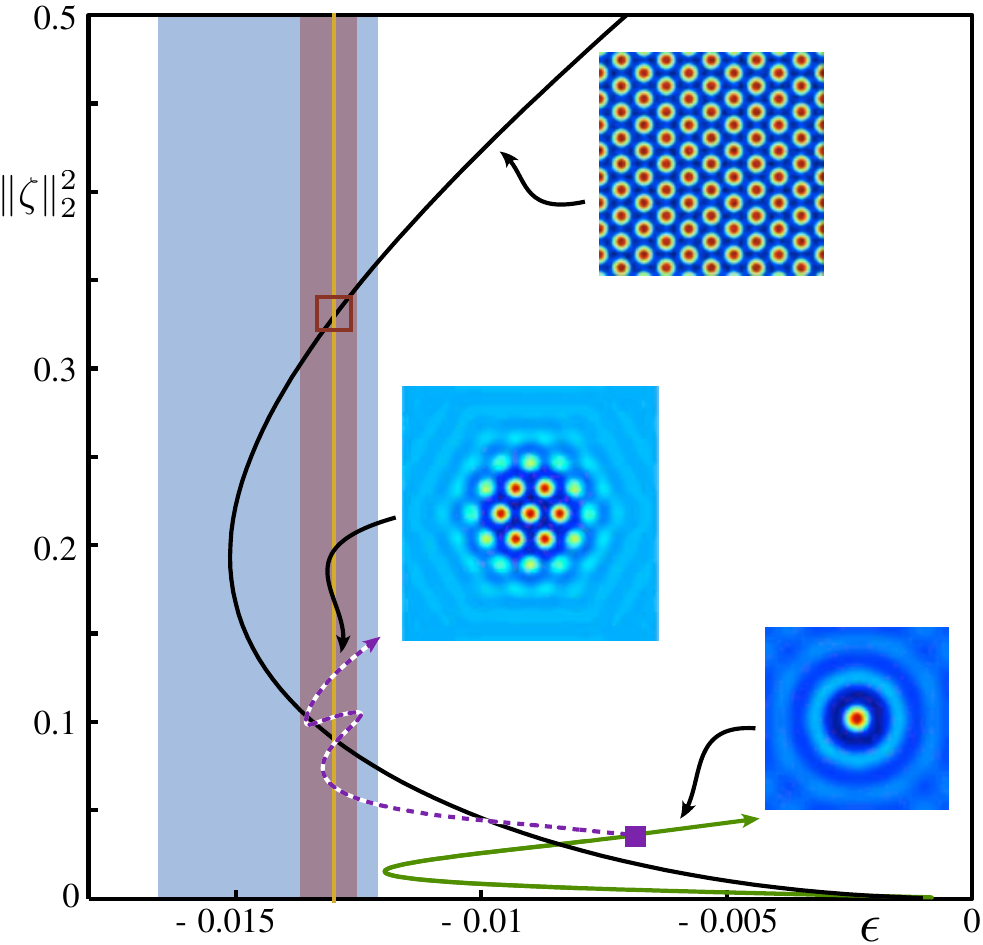}
\caption{Emergence of a fully localized hexagon patch (dashed purple) bifurcating from a single spike (green) with $\mu_r=2$. We also plot the snaking regions of the $\langle10\rangle$- (blue) and $\langle11\rangle$-planar fronts (red).\label{f:patches}}
\end{figure}

\subsection{Hexagon patches}\label{ss:hexagon.patches}
We turn now from the planar fronts to the hexagonal patches as seen in the experiment.
The numerical results for $\mu_r=2$ displayed in figure~\ref{f:patches} show
that radial spots bifurcate subcritically at onset from the flat state and undergo a subsequent fold. Along the upper branch of the spot there is a $\mathbb{D}_6$-bifurcation from which a hexagon patch emerges that undergoes a series of folds yielding larger and larger patches. We terminate the branch early as the snaking branch becomes highly intricate further up the branch (cf. \citet[Figure 25]{lloyd2008}). However, we expect this branch to eventually terminate near the fold of the domain-covering hexagons. This patch also snakes about around the $\miller{10}$- and $\miller{11}$-hexagon fronts (indicated by the shaded regions). We observe that the snaking occurs around the hexagon fronts since large patches are effectively made up of combinations of hexagon fronts's sides.
The hexagon patch displayed in the central inset of figure~\ref{f:patches} occurs after the third fold of the hexagon-patch-branch, and comprises seven fully developed spikes together with an outer shell of
12 lower humps. It resembles the experimental patch presented in figure~\ref{f:7patch}, which is also part of the $\tau_2 = 7.5\,\mathrm{s}$ series in figure~\ref{f:exp_step}.

\subsection{Rhomboid patches}\label{ss:rhomboid.patches}
As seen in the experimental patterns shown in figure~\ref{f:exp_loc}, the localisation of the patches is not necessarily centred around a spike. The patches may equally well organize around a valley in-between two spikes. These structures have been called localised rhomboids~\citep{lloyd2008}. As shown in figure~\ref{f:rhomboids}, we are able to find these structures in the same parameter region as the localised hexagon patches. There are some crucial differences for these structures to the localised hexagon patches with a spike at the centre of the patch. The first difference is that rhomboid patches do not emerge from a bifurcation of a single spike but from a double spike configuration that bifurcates from the flat state. Figure~\ref{f:rhomboids} explains how patches are grown by first trying to complete a $\miller{10}$-front side by first adding cells in the middle of each of these sides. As the patches get larger, we see the snaking becomes more confined to the $\miller{10}$- and $\miller{11}$-front snaking regions. In particular, when a patch has only  $\miller{10}$-front sides, then the snaking diagram makes its largest excursion by travelling from the left-most $\miller{11}$-front limit to the right-most $\miller{10}$. All other types of patches are confined to the $\miller{11}$-front region. We note that the existence regions of the various configurations are not the same.
\begin{figure}
\centering
\includegraphics[width=1\linewidth]{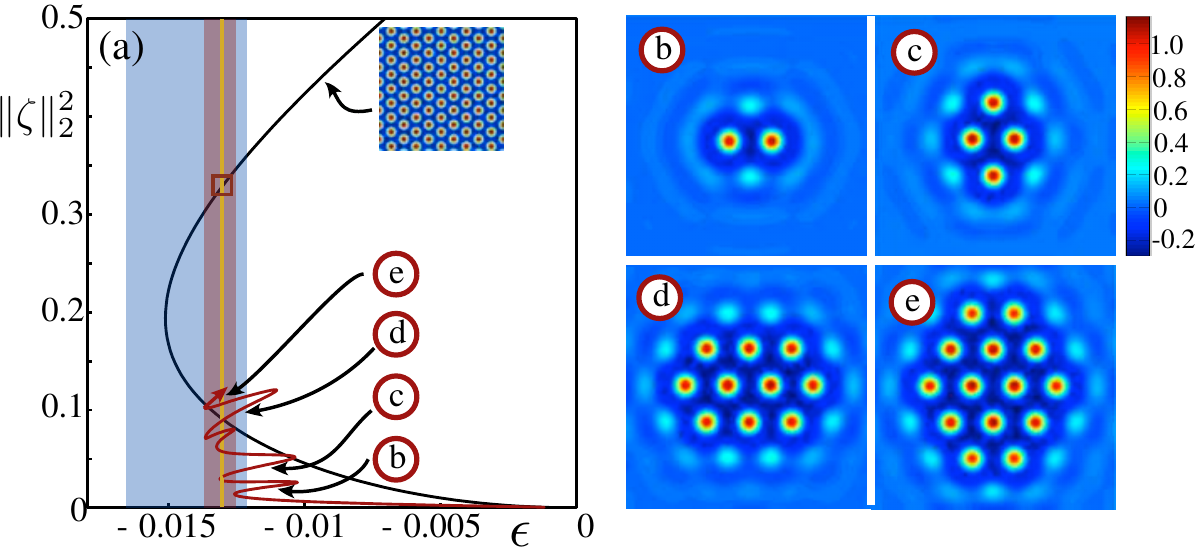}
\caption{Emergence of fully localized rhomboid patches bifurcating from the flat state with $\mu_r=2$. We also plot the snaking regions of the $\langle10\rangle$- (blue) and $\langle11\rangle$-planar fronts (red). \label{f:rhomboids}}
\end{figure}

We shall briefly comment on stability of the branch shown in figure~\ref{f:rhomboids}. It is not possible to compute the stability of the branches directly from the numerics since we have not written down an equation for the evolution of the fluid velocity. However, we expect the branch emanating from $\epsilon=0$ to initially be unstable since the bifurcation is subcritical. We anticipate that the branch then re-stabilises at the first fold to yield a stable two-spike patch and between each subsequent fold, the branch will destabilise (due to an eigenvalue crossing the imaginary axis at the fold) and then re-stabilise; see~\citep{beck2009}. Near each fold, we anticipate that there are symmetry-breaking bifurcations leading to non-$\mathbb{D}_2$ symmetric patches.

\begin{figure}
\centering
\includegraphics[width=0.6\linewidth]{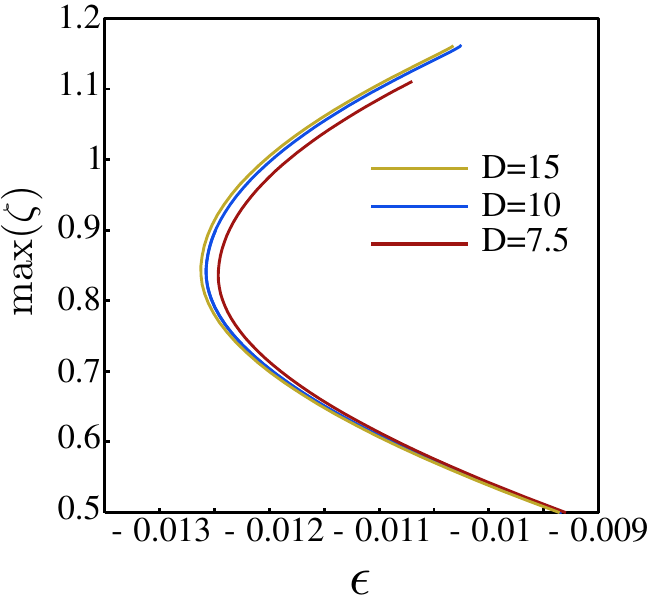}
\caption{The height of the rhomboid patches for $D=7.5,10,$ and $15$ for $\mu_r=2$ up to the second fold. We observe that as we decrease $D$ the height of the patches also decreases but that for $D\geq10$ we observe little change in the height.\label{f:height_D}}
\end{figure}

We eventually  investigate the effect of changing the depth $D$ on the localised rhomboid patches. In figure\,\ref{f:height_D}, we plot the maximum height of the patches for $D=7.5,10,$ and $15$.
For clarity we are stopping the bifurcation curves before the second fold.
As the depth increases the snaking diagrams are shifted to the left and the maximum height of the patches increases as well. It becomes clear that the existence regions of the localized patches are affected by the depth of the fluid. This is believed to be caused by the Neumann boundary conditions at the bottom (which insist that the magnetic field decays to the applied magnetic field here). In a shallow layer, that constraint also reduces the magnetic field within the tip of the spikes, and thus their maximum height. 

\section{Comparison between experimental and numerical results}\label{s:comparison}
We now turn to comparing our experimental and numerical results. As stated in \S\ref{s:gov.eqs} our model is based on a linear magnetisation law, whereas we presented in \S \ref{s:ferro_properties} the nonlinear $M(H_\mathrm{F})$ of the real ferrofluid, with an effective permeability ranging from $\mu_\mathrm{eff}=4.57$ at $B=0\,\mathrm{mT}$ down to 3.2 at $B_c=10.8\,\mathrm{mT}$.
Therefore a direct matching of all experimental and numerical results can not be expected. For $\mu_r\rightarrow1$, the linear magnetisation law becomes a more accurate approximation of the nonlinear magnetisation law and the model predicts homoclinic snaking should persist, albeit with an exponentially small width the snaking exists in.

In order to provide a comparison with the experimental results in \S\ref{s:exp_results}, we instead chose a magnetic permeability $\mu_r=\mathrm{const}$ for our model that captures the qualitative features of the experiment. In a first attempt we chose $\mu_r$ to be the effective permeability at the critical point $\mu_\mathrm{eff}(H_c) = 3.2$ (cf. figure\,\ref{f:magnetisation}\,b) for the model: $\mu_r = 3.2$.
In figure \ref{f:bif_mu_3_2}\,(a), we path-follow the two-spike patch, as depicted in figure  \ref{f:bif_mu_3_2}\,(b), bifurcating from the flat state to the upper branch.
\begin{figure}
	\centering
	\includegraphics[width=1\linewidth]{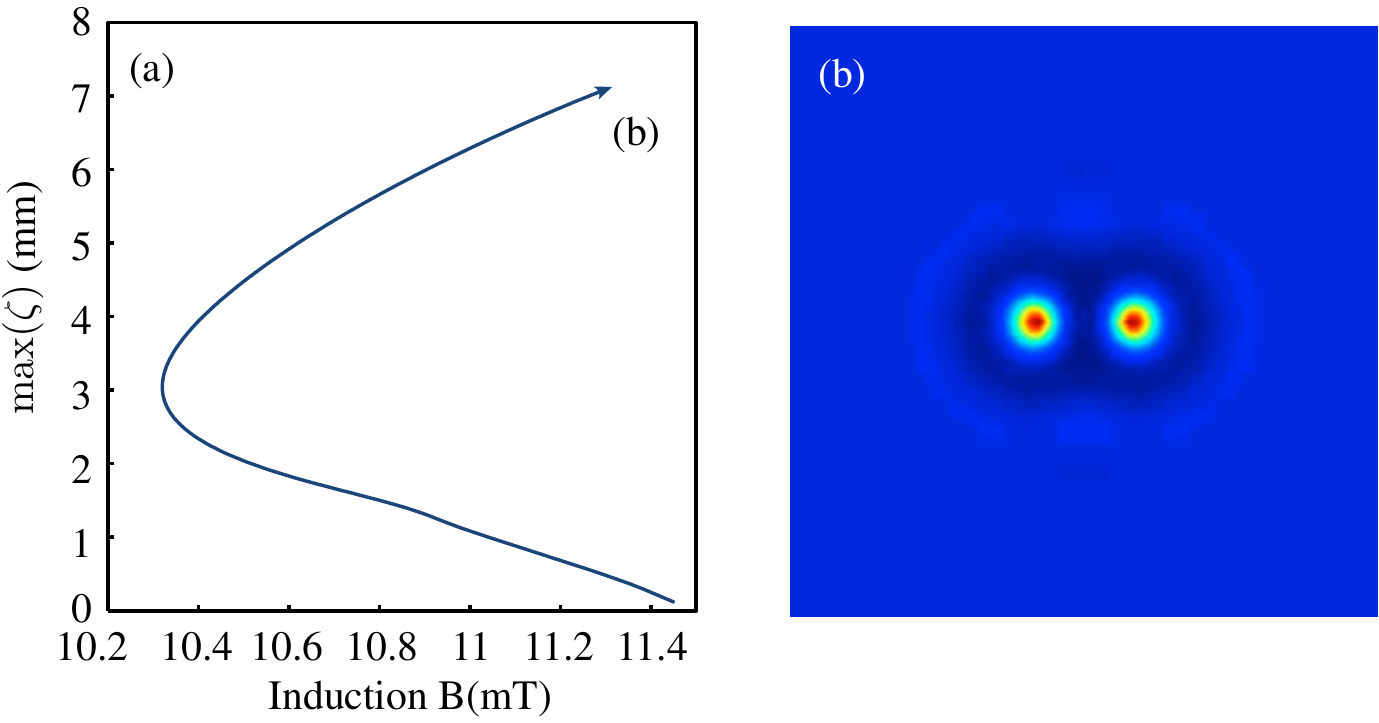}
	\caption{(a) Numerical bifurcation diagram for $\mu_r=3.2$ for a two-spike configuration, depicted in (b), that has bifurcated off the flat state. The state in (b) was computed for $B=11.3$\,mT. We terminated the branch early.
\label{f:bif_mu_3_2}}
\end{figure}
As seen from figure \ref{f:bif_mu_3_2}\,(a), for $\mu_r=3.2$ localised structures exist from $B=10.3$\,mT to $11.5$\,mT which is a significantly larger range (about 200\%) than that observed in figure \ref{f:exp_step}. Moreover the height of the spikes is about 8\% larger than that observed experimentally.

In a second attempt, we select a smaller value of magnetic permeability, namely $\mu_r=2$. Here we find that while the critical value of the induction is far from that experimentally observed, the width of the bistability region between the flat state and the domain-covering hexagons is very close (approximately $0.25\,\mathrm{mT}$ in both experiments and numerics); see figure~\ref{f:rhomboids_height}. Furthermore, we find from the numerics at the fold of the domain-covering hexagons the predicted height of the hexagons is 1.2\,mm (from figure~\ref{f:exp_protocol} we find the measured height of 1.3\,mm) and at onset the height of the stable hexagons is predicted to be 2.4\,mm (and from figure~\ref{f:exp_protocol} we find the measured height of 2.4\,mm) i.e., in the bistability region the numerics are within 5\% error of that experimentally observed. Hence, it is possible that phenomenologically $\mu_r=2$ is the appropriate value to take in the linear magnetisation law for the numerics for localised hexagon patches.

\begin{figure}
\centering
\includegraphics[width=0.5\linewidth]{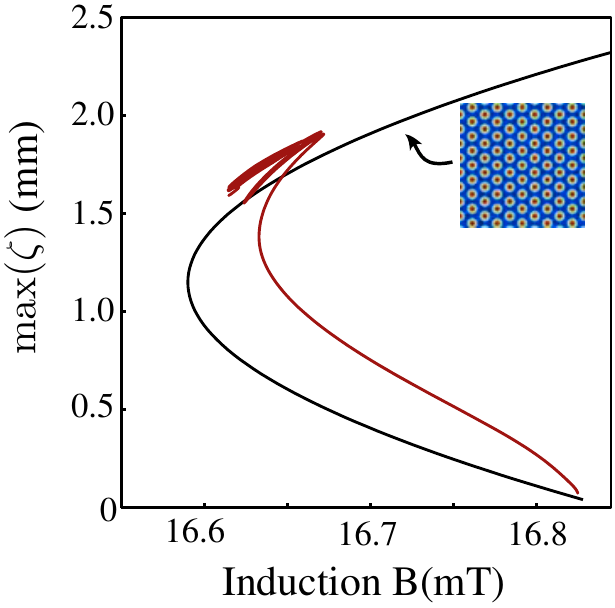}
\caption{Maximum height of the domain covering pattern (black line) computed numerically for $\mu_r=2$ at the critical wavenumber $k_\mathrm{c}\approx608.94\,\mathrm{m^{-1}}$ versus the magnetic induction. The maximum height of the rhomboid patches is marked by a brown line.
\label{f:rhomboids_height}}
\end{figure}

The heights of the spikes in the patches are found numerically to be highest at the core of the patch and decrease as the spikes get closer to the edge of the patch. As shown in Figure~\ref{f:rhomboids_height}(a), the maximum height of the rhomboid patches is slightly larger than the domain-covering hexagons similar to that seen in the experiments.

The numerical observation of various co-existing hexagon patches in figure \ref{f:patches}, in the bistability region of the flat state and the domain-covering hexagons, agrees well with the experimental results shown in figure\,\ref{f:exp_step} where for the same $B_3$-values several different patches were observed. In particular, we are able to find numerically patches (figure~\ref{f:rhomboids} (a), (b), (c) \& (d)) that look to be identical to those seen in figure\,\ref{f:exp_loc}(b), (d), (f) and (i). Our numerical methods imposed that the patches should have $\mathbb{D}_2$ symmetry hence we were unable to locate any of the other patches seen in figure\,\ref{f:exp_loc}. However, due to the existence of a Maxwell point near-by, we believe hexagon spikes/cells can be added to the patches to break the $\mathbb{D}_2$ symmetry provided you are in the $\miller{11}$-hexagon front snaking region. Hence, we have a possible explanation for the existence of all the various types of patches observed experimentally.

Perhaps the largest difference between the numerical and experimental results is that
the sequence of plateaus for fixed $\tau_2$ suggests that one can climb up the snakes for increasing $B_3$, i.e. the snakes are not vertical as in figure\,\ref{f:patches} or \ref{f:rhomboids} but slanted, as found by \citet{firth2007} and \citet{dawes2008} for a conserved quantity. Slanted snaking may emerge in the presence of a nonlinear magnetisation law and finite-domain effects and
could provide an explanation for the emergence of larger and larger patches seen in figure\,\ref{f:rhomboids}.

\section{Conclusion and Outlook}\label{s:conclusion}
Utilizing a magnetic pulse sequence we prepare a ferrofluidic layer to be situated next to the unstable branch of the Rosensweig instability. There localized hexagon patches emerge spontaneously. The measured sequence of patches indicates homoclinic snaking. This is corroborated by numerics where we investigate the Young-Laplace equation coupled to the Maxwell equations.
Homoclinic snaking is demonstrated for two types of planar hexagon fronts as well as hexagon and rhomboid patches. The experimentally observed sequence of patterns can be interpreted as an imperfect mixture of those prototype scenarios. The numerically unveiled Maxwell point of the hexagons is providing an energy argument for those various regular and irregular patches observed since near the Maxwell point the energy of a single hexagon spike is the same as the flat state and there is no energy penalty in adding/removing hexagon spikes.
The width of the experimentally observed plateaus is about 0.5\% of the control parameter,
which is comparable to the snaking interval of the numerical prototypes. Moreover, the height of the patches in experiment and numerics agrees within 10\%.
Therefore, we believe that our model captures all the most important phenomena of the system. In particular at the relative magnetic permeability tends to unity the validity of our model increases. It thus unveils an {\it organising centre} for the emergence of localised hexagon patches seen in the experiment.

It still remains to explore how various stable configurations of patches connect via unstable branches both experimentally (by quasi-statically varying the magnetic induction {$B$ once a patch has been generated)} and theoretically. The discrepancy between the experimental indications for slanted snaking, and the vertical snaking obtained  numerically may be solved by incorporating a nonlinear magnetisation law, finite-size and shallow layer effects or by carrying out more experiments to determine the precise extent of the plateaus.

Theoretically, there are now some very interesting directions one could take. With the existence of an energy functional for the full system that incorporates the boundary conditions one could look at applying variational and spatial-dynamical techniques that have been developed in the water-wave literature; see for instance~\citep{groves2008,buffoni2013}.

\section*{Acknowledgements}
The authors confirm that data underlying the findings are available without restriction. Details of the data and how to request access are available from the University of Surrey publications repository: \url{http://epubs.surrey.ac.uk/id/eprint/807861} 

DJBL acknowledges support from an EPSRC grant (Nucleation of Ferrosolitons and Ferropatterns) EP/H05040X/1 and many fruitful discussions with Mark Groves. Moreover the authors thank Achim Beetz and Robin Maretzki for recording the high density range picture in figure\,1, and Klaus Oetter for assistance in constructing the experimental setup. In memory of Rudolf Friedrich.

\section*{Appendix: Validation of the numerical scheme}
We present an overview of the validation results of the numerical scheme. We first plot a convergence test for the domain covering hexagons for $(\epsilon,\mu_r)=(0,2)$ on the domain $(x,y)\in[-2\pi,2\pi]\times[-2\pi/\sqrt{3},2\pi/\sqrt{3}]$ and $D=10$. In figure\,\ref{f:convergence}, we plot the  the relative error (relative to a hexagon spike computed with $N_x=N_y=40$) as $N_x=N_y=N$ is varied and then as $Nz$ is varied. Since we are using pseudo-spectral methods, we see rapid (geometric) convergence with few mesh points. For our numerical results we have a relative error of $10^{-3}$ and the fold points are computed to a relative error of $10^{-4}$.
\begin{figure}
\centering
\includegraphics[width=0.8\linewidth]{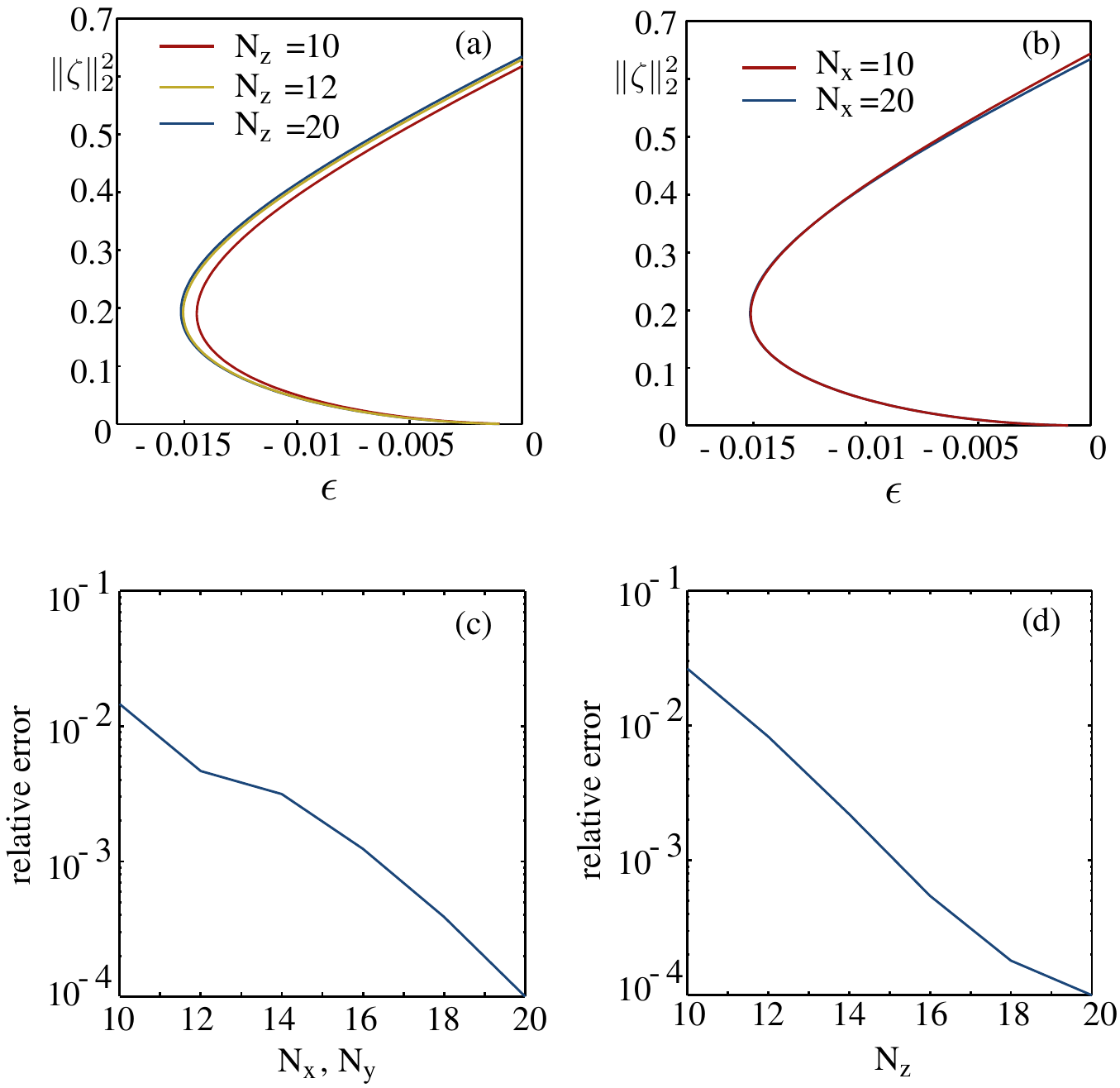}
\caption{(a) Bifurcation diagram of the domain covering hexagons for $\mu_r=2$ on the domain $(x,y)\in[-2\pi,2\pi]\times[-2\pi/\sqrt{3},2\pi/\sqrt{3}]$ with $N_x=N_y=20,D=10$ and varying $N_z$. (b) Bifurcation diagram of the domain covering hexagons for $\mu_r=2$ on the domain $(x,y)\in[-2\pi,2\pi]\times[-2\pi/\sqrt{3},2\pi/\sqrt{3}]$ with $N_z=20$ and varying $N_x$ ($N_y=N_x$). (c) Relative error on a semi-log scale of the L2 norm of the interface $\|\zeta\|_2^2$ for the domain covering hexagons  on the upper branch at $(\epsilon,\mu_r)=(0,2)$ with $N_z=20,D=10$ and varying $N_x=N_y$ with respect to the hexagon computed for $N_z=22,N_x=N_y=22$. (d) Relative error on a semi-log scale of the L2 norm of the interface $\|\zeta\|_2^2$ for the domain covering hexagons  on the upper branch at $(\epsilon,\mu_r)=(0,2)$ with $N_x=N_y=20,D=10$ and varying $N_z$ with respect to the hexagon computed for $N_z=22,N_x=N_y=22$. From both panels (c) \& (d) we observe rapid (geometric) convergence due to the spectral accuracy of the numerical pseudo-spectral methods.\label{f:convergence}}
\end{figure}

\begin{figure}
\centering
\includegraphics[width=0.8\linewidth]{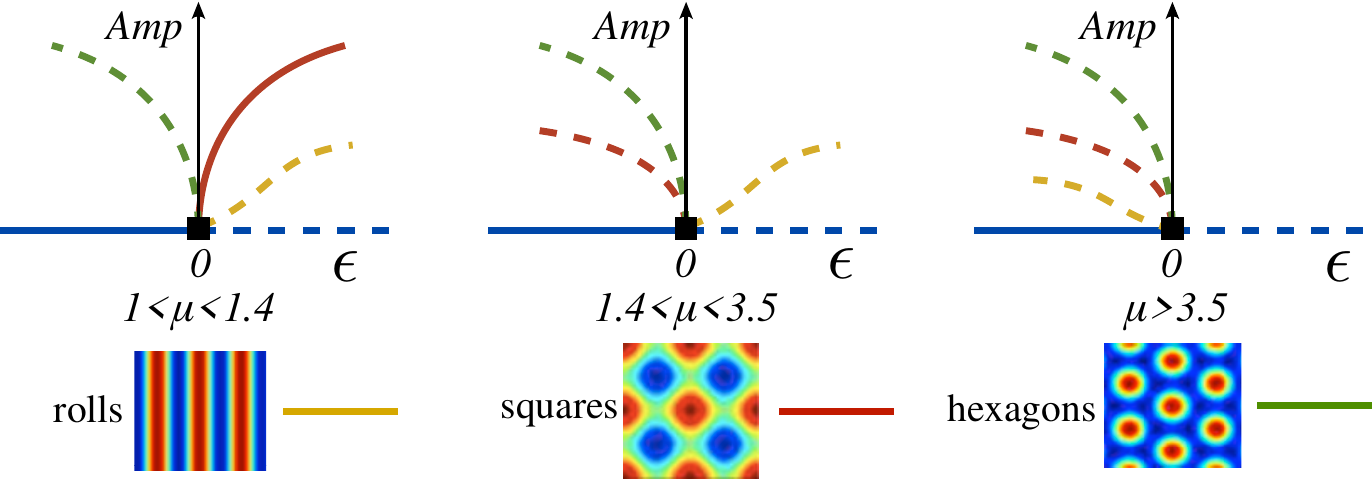}
\caption{Normal form results of \cite{silber1988} showing the changes in sub- and super-criticality of the bifurcations for rolls, squares and up-hexagons i.e hexagons whose maximum amplitude is positive.\label{f:silber}}
\end{figure}
In order to test that we have correctly captured the nonlinearity in the system, we compare our results with the normal form analysis of \citet{silber1988} shown in figure~\ref{f:silber}. Since the analysis is for small amplitude we only compare the bifurcation transitions for each of the rolls and squares from sub- to supercritical. Here we find excellent agreement with our numerics giving us confidence in our results; shown in figure~\ref{f:num_valid_2} where we plot the existence regions of the rolls, squares and hexagons for $\epsilon<0$. The final validation we did was to find that the Maxwell point computed for $\mu_r=1.9$ lies in the middle of the snaking region of both the $\miller{10}$- and $\miller{11}$-fronts in figure~\ref{f:planar_fronts} as predicted by~\citep{woods1999,coullet2000}. This demonstrates that the discretisation of (\ref{e:sys_b},\ref{e:sys_c}) and~(\ref{e:comp_a}) is consistent with the energy/hamiltonian formalisation~(\ref{e:energy}) and (\ref{e:ham}) and we are solving the correct equations.
\begin{figure}
\centering
\includegraphics[width=0.5\linewidth]{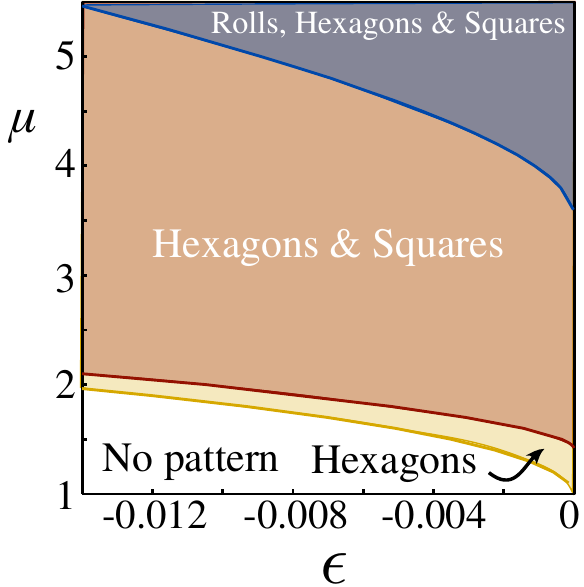}
\caption{Numerical computation of existence regions of domain-covering rolls, squares and hexagons for $\epsilon<0$. We observe that all three of the domain-covering patterns bifurcate subcritically in the regions where the normal form theory of \cite{silber1988} predicts. \label{f:num_valid_2}}
\end{figure}

\bibliographystyle{jfm}
\bibliography{localised_pattern_bib_2015_01_17}

\begin{thebibliography}{56}
\expandafter\ifx\csname natexlab\endcsname\relax\def\natexlab#1{#1}\fi

\bibitem[Abshagen {\em et~al.\/}(2010)Abshagen, Heise, Pfister \&
  Mullin]{abshagen2010}
{\sc Abshagen, J., Heise, M., Pfister, G. \& Mullin, T.} 2010 Multiple
  localized states in centrifugally stable rotating flow. {\em Phys. Fluids\/}
  {\bf 22}, 021702.

\bibitem[Akhmediev \& Ankiewicz(2005)]{akhmediev2005}
{\sc Akhmediev, N. \& Ankiewicz, A.}, ed. 2005 {\em Dissipative Solitons,
  Lectures Notes in Physics\/}, 1st edn. Springer.

\bibitem[Ashcroft \& Mermin(1976)]{ashcroft1976}
{\sc Ashcroft, Neil~W. \& Mermin, N.~David} 1976 {\em Solid {S}tate
  {P}hysics\/}. New York: Harcourt.

\bibitem[Barbay {\em et~al.\/}(2008)Barbay, Hachair, Elsass, Sagnes \&
  Kuszelewicz]{barbay2008hss}
{\sc Barbay, S., Hachair, X., Elsass, T., Sagnes, I. \& Kuszelewicz, R.} 2008
  Homoclinic snaking in a semiconductor-based optical system. {\em Phys. Rev.
  Lett.\/} {\bf 101}~(25), 253902.

\bibitem[Batiste {\em et~al.\/}(2006)Batiste, Knobloch, Alonso \&
  Mercader]{batiste2006}
{\sc Batiste, Oriol, Knobloch, Edgar, Alonso, Arantxa \& Mercader, Isabel} 2006
  Spatially localized binary-fluid convection. {\em J. Fluid Mech.\/} {\bf
  560}, 149--158.

\bibitem[Beck {\em et~al.\/}(2009)Beck, Knobloch, Lloyd, Sandstede \&
  Wagenknecht]{beck2009}
{\sc Beck, M., Knobloch, J., Lloyd, D.J.B., Sandstede, B. \& Wagenknecht, T.}
  2009 Snakes, ladders, and isolas of localised patterns. {\em SIAM J. Math.
  Anal.\/} {\bf 41}, 936--972.

\bibitem[Bohlius {\em et~al.\/}(2011)Bohlius, Brand \& Pleiner]{bohlius2011aer}
{\sc Bohlius, Stefan, Brand, Helmut~R. \& Pleiner, Harald} 2011 Amplitude
  equation for the {Rosensweig} instability. {\em Prog. Theor. Phys.\/} {\bf
  125}, 1--46.

\bibitem[Bohlius {\em et~al.\/}(2006)Bohlius, Pleiner \& Brand]{bohlius2006}
{\sc Bohlius, S., Pleiner, H. \& Brand, Helmut~R.} 2006 {Pattern Formation in
  Ferrogels: Analysis of the Rosensweig Instability Using the Energy Method}.
  {\em J. Phys.: Condens. Matter\/} {\bf 18}~(38), 2671--2684.

\bibitem[Buffoni {\em et~al.\/}(2013)Buffoni, Groves, Sun \&
  Wahl{\'e}n]{buffoni2013}
{\sc Buffoni, B., Groves, M.~D., Sun, S.~M. \& Wahl{\'e}n, E.} 2013 Existence
  and conditional energetic stability of three-dimensional fully localised
  solitary gravity-capillary water waves. {\em J. Differential Equations\/}
  {\bf 254}~(3), 1006--1096.

\bibitem[Cao \& Ding(2014)]{cao2014}
{\sc Cao, Y \& Ding, Z.J.} 2014 Formation of hexagonal pattern of ferrofluid in
  magnetic field. {\em {J. Magn. Mag. Mater.}\/} {\bf 355}, 93--99.

\bibitem[Chantry \& Kerswell(2014)]{chantry2014}
{\sc Chantry, M. \& Kerswell, R.~R.} 2014 {Localization in a spanwise-extended
  model of plane Couette flow}. Preprint.

\bibitem[Coullet {\em et~al.\/}(2000)Coullet, Riera \& Tresser]{coullet2000}
{\sc Coullet, P., Riera, C. \& Tresser, C.} 2000 Stable static localized
  structures in one dimension. {\em Phys. Rev. Lett.\/} {\bf 84}~(14),
  3069--3072.

\bibitem[Cowley \& Rosensweig(1967)]{cowley1967}
{\sc Cowley, M.~D. \& Rosensweig, R.~E.} 1967 The interfacial stability of a
  ferromagnetic fluid. {\em J. Fluid Mech.\/} {\bf 30}~(04), 671--688.

\bibitem[Dawes(2008)]{dawes2008}
{\sc Dawes, J. H.~P.} 2008 Localised pattern formation with a large scale mode:
  slanted snaking. {\em SIAM J. Appl. Dyn. Syst.\/} {\bf 7}~(1), 186--206.

\bibitem[Dawes(2010)]{dawes2010}
{\sc Dawes, J. H.~P.} 2010 The emergence of a coherent structure for coherent
  structures: localized states in nonlinear systems. {\em Philos. Trans. R.
  Soc. Lond. Ser. A Math. Phys. Eng. Sci.\/} {\bf 368}~(1924), 3519--3534.

\bibitem[Dean {\em et~al.\/}(2014)Dean, Matthews, Cox \& King]{dean2014}
{\sc Dean, A., Matthews, P.~C., Cox, S.~M. \& King, J.} 2014
  {Orientation-dependent pinning and homoclinic snaking on a planar lattice}.
  Preprint.

\bibitem[Descalzi {\em et~al.\/}(2011)Descalzi, Clerc, Residori \&
  Assanto]{descalzi2011}
{\sc Descalzi, Orazio, Clerc, Marcel, Residori, Stefania \& Assanto, Gaetano},
  ed. 2011 {\em {Localized States in Physics: Solitons and Patterns}\/}.
  Springer.

\bibitem[Escaff \& Descalzi(2009)]{escaff2009}
{\sc Escaff, D. \& Descalzi, O.} 2009 Shape and size effects in localized
  hexagonal patterns. {\em Internat. J. Bifur. Chaos Appl. Sci. Engrg.\/} {\bf
  19}, 2727--2743.

\bibitem[Firth {\em et~al.\/}(2007)Firth, Columbo \& Scroggie]{firth2007}
{\sc Firth, W.~J., Columbo, L. \& Scroggie, A.~J.} 2007 {Proposed Resolution of
  Theory-Experiment Discrepancy in Homoclinic Snaking}. {\em Phys. Rev.
  Lett.\/} {\bf 99}, 104503.

\bibitem[Friedrichs(2002)]{friedrichs2002}
{\sc Friedrichs, Rene} 2002 Low symmetry patterns on magnetic fluids. {\em
  Phys. Rev. E\/} {\bf 66}, 066215~1--7.

\bibitem[Friedrichs \& Engel(2001)]{friedrichs2001}
{\sc Friedrichs, R \& Engel, A} 2001 Pattern and wave number selection in
  magnetic fluids. {\em Phys. Rev. E\/} {\bf 64}, 021406.

\bibitem[Fr{\"o}hlich(1881)]{Froehlich1881}
{\sc Fr{\"o}hlich, O.} 1881 Investigations of dynamoelcetric machines and
  electric power transmission and theoretical conclusions therefrom. {\em
  Elektrotech. Z\/} {\bf 2}, 134 -- 141.

\bibitem[Gailitis(1977)]{gailitis1977}
{\sc Gailitis, A.} 1977 {Formation of the hexagonal pattern on the surface of a
  ferromagneticfluid in an applied magnetic field}. {\em J. Fluid Mech.\/} {\bf
  82}~(3), 401--413.

\bibitem[Gollwitzer {\em et~al.\/}(2007)Gollwitzer, Matthies, Richter, Rehberg
  \& Tobiska]{gollwitzer2007}
{\sc Gollwitzer, C., Matthies, G., Richter, R., Rehberg, I. \& Tobiska, L.}
  2007 {The surface topography of a magnetic fluid: a quantitative comparison
  between experiment and numerical simulation}. {\em J. Fluid Mech.\/} {\bf
  571}, 455--474.

\bibitem[Gollwitzer {\em et~al.\/}(2010)Gollwitzer, Rehberg \&
  Richter]{gollwitzer2010}
{\sc Gollwitzer, C., Rehberg, I. \& Richter, R.} 2010 From phase space
  representation to amplitude equations in a pattern-forming experiment. {\em
  {New J. Phys.}\/} {\bf 12}, 093037.

\bibitem[Groves {\em et~al.\/}(2015)Groves, Lloyd \& Stylianou]{groves2015}
{\sc Groves, M.~D., Lloyd, D. J.~B. \& Stylianou, A.} 2015 {Pattern formation
  on the free surface of a ferrofluid: Spatial dynamics and homoclinic
  bifurcation}. In preparation.

\bibitem[Groves \& Sun(2008)]{groves2008}
{\sc Groves, M.~D. \& Sun, S.-M.} 2008 Fully localised solitary-wave solutions
  of the three-dimensional gravity-capillary water-wave problem. {\em Arch.
  Ration. Mech. Anal.\/} {\bf 188}~(1), 1--91.

\bibitem[Groves \& Toland(1997)]{groves1997}
{\sc Groves, Mark~D. \& Toland, John~F.} 1997 On variational formulations for
  steady water waves. {\em Arch. Rational Mech. Anal.\/} {\bf 137}~(3),
  203--226.

\bibitem[Haudin {\em et~al.\/}(2011)Haudin, Rojas, Bortolozzo, Residori \&
  Clerc]{haudin2011}
{\sc Haudin, F., Rojas, R.~G., Bortolozzo, U., Residori, S. \& Clerc, M.~G.}
  2011 Homoclinic snaking of localized patterns in a spatially forced system.
  {\em Phys. Rev. Lett.\/} {\bf 107}, 264101.

\bibitem[Kelley(2003)]{kelley2003}
{\sc Kelley, C.~T.} 2003 {\em {Solving Nonlinear Equations with Newton's Method
  }\/}. {\em Fundamental Algorithms for Numerical Calculations\/} 1. SIAM,
  Philadelphia.

\bibitem[Knieling {\em et~al.\/}(2010)Knieling, Rehberg \&
  Richter]{knieling2010}
{\sc Knieling, H., Rehberg, I. \& Richter, R.} 2010 The growth of localized
  states on the surface of magnetic fluids. {\em Physics Procedia\/} {\bf 9},
  199.

\bibitem[Knobloch(2008)]{knobloch2008}
{\sc Knobloch, E.} 2008 Spatially localized structures in dissipative systems:
  open problems. {\em Nonlinearity\/} {\bf 21}~(4), T45--T60.

\bibitem[Knobloch(2015)]{knobloch2015}
{\sc Knobloch, E.} 2015 {Spatial Localization in Dissipative Systems}. {\em
  Annu. Rev. Condens. Matter Phys.\/} {\bf 6}, 325--59.

\bibitem[Kozyreff \& Chapman(2013)]{kozyreff2013}
{\sc Kozyreff, G. \& Chapman, S.~J.} 2013 {Analytical Results for Front Pinning
  between an Hexagonal Pattern and a Uniform State in Pattern-Formation
  Systems}. {\em Phys. Rev. Lett.\/} {\bf 111}, 054501.

\bibitem[Krauskopf {\em et~al.\/}(2007)Krauskopf, Osinga \&
  Galan-Vioque]{krauskopf2007}
{\sc Krauskopf, B., Osinga, H.~M. \& Galan-Vioque, J.}, ed. 2007 {\em Numerical
  Continuation Methods for Dynamical Systems\/}. Springer.

\bibitem[Lavrova {\em et~al.\/}(2008)Lavrova, Matthies \& Tobiska]{lavrova2008}
{\sc Lavrova, O., Matthies, G. \& Tobiska, L.} 2008 {Numerical study of
  soliton-like surface configurations on a magnetic fluid layer in the
  Rosensweig instability}. {\em Communications in Nonlinear Science and
  Numerical Simulation\/} {\bf 13}, 1302--1310.

\bibitem[Lloyd \& O'Farrell(2013)]{lloyd2013}
{\sc Lloyd, David J.~B. \& O'Farrell, Hayley} 2013 On localised hotspots of an
  urban crime model. {\em Phys. D\/} {\bf 253}, 23--39.

\bibitem[Lloyd {\em et~al.\/}(2008)Lloyd, Sandstede, Avitabile \&
  Champneys]{lloyd2008}
{\sc Lloyd, D. J.~B., Sandstede, B., Avitabile, D. \& Champneys, A.~R.} 2008
  Localized hexagon patterns of the planar {S}wift--{H}ohenberg equation. {\em
  SIAM J. Appl. Dynam. Syst.\/} {\bf 7}, 1049--1100.

\bibitem[Mercader {\em et~al.\/}(2013)Mercader, Batiste, Alonso \&
  Knobloch]{mercader2013}
{\sc Mercader, I., Batiste, O., Alonso, A. \& Knobloch, E.} 2013 Travelling
  convectons in binary fluid convection. {\em J. Fluid Mech.\/} {\bf 722},
  240--266.

\bibitem[Moses {\em et~al.\/}(1987)Moses, Fineberg \& Steinberg]{moses1987}
{\sc Moses, Elisha, Fineberg, Jay \& Steinberg, Victor} 1987 Multistability and
  confined traveling-wave patterns in a convecting binary mixture. {\em Phys.
  Rev. A\/} {\bf 35}, 2757--2760.

\bibitem[Pomeau(1986)]{pomeau1986}
{\sc Pomeau, Y.} 1986 Front motion, metastability, and subcritical bifurcations
  in hydrodynamics. {\em Physica~D\/} {\bf 23}~(1-3), 3--11.

\bibitem[Pringle {\em et~al.\/}(2014)Pringle, Willis \& Kerswell]{pringle2014}
{\sc Pringle, Chris~C.T., Willis, Ashley~P. \& Kerswell, Rich~R.} 2014 Fully
  localised nonlinear energy growth optimals in pipe flow. Preprint.

\bibitem[Purwins {\em et~al.\/}(2010)Purwins, B{\"{o}}deker \&
  Amiranashvili]{purwins2010}
{\sc Purwins, H.-G., B{\"{o}}deker, H.~U. \& Amiranashvili, Sh.} 2010
  Dissipative solitons. {\em Advances in Physics\/} {\bf 59}~(5).

\bibitem[Rault(2000)]{rault2000}
{\sc Rault, J.} 2000 Origin of the {Vogel-Fulcher-Tammann} law in glass-forming
  materials: the $\alpha-\beta$ bifurcation. {\em J. Non-Cryst. Solids\/} {\bf
  271}, 177.

\bibitem[Richter \& Barashenkov(2005)]{richter2005}
{\sc Richter, R. \& Barashenkov, I.~V.} 2005 Two-dimensional solitons on the
  surface of magnetic fluids. {\em Phys. Rev. Lett.\/} {\bf 94}, 184503.

\bibitem[Richter \& Bl{\"a}sing(2001)]{richter2001}
{\sc Richter, R. \& Bl{\"a}sing, J.} 2001 Measuring surface deformations in
  magnetic fluid by radioscopy. {\em Rev. Sci. Instrum.\/} {\bf 72},
  1729--1733.

\bibitem[Rosensweig(1985)]{rosensweig1985}
{\sc Rosensweig, R.~E.} 1985 Directions in ferrohydrodynamics. {\em Journal of
  Applied Physics\/} {\bf 57}~(8), 4259--4264.

\bibitem[Saad(1996)]{saad1996}
{\sc Saad, Yousef} 1996 {\em Iterative Methods for Sparse Linear Systems,\/},
  chap. Chapter 10 - Preconditioning Techniques. PWS Publishing Company.

\bibitem[Schneider {\em et~al.\/}(2010)Schneider, Gibson \&
  Burke]{schneider2010}
{\sc Schneider, T.~M., Gibson, J.~F. \& Burke, John} 2010 Snakes and ladders:
  localized solutions of plane {C}ouette flow. {\em Phys. Rev. Lett.\/} {\bf
  104}, 104501.

\bibitem[Silber \& Knobloch(1988)]{silber1988}
{\sc Silber, Mary \& Knobloch, Edgar} 1988 Pattern selection in ferrofluids.
  {\em Phys. D\/} {\bf 30}~(1-2), 83--98.

\bibitem[Thompson(2015)]{thompson2015}
{\sc Thompson, J. Michael~T.} 2015 Advances in shell buckling: Theory and
  experiments. {\em Int. J. Bifurcation and Chaos\/} {\bf 25}~(1), 1530001--1.

\bibitem[Twombly \& Thomas(1983)]{twombly1983}
{\sc Twombly, Evan~Eugene \& Thomas, J.~W.} 1983 Bifurcating instability of the
  free surface of a ferrofluid. {\em SIAM J. Math. Anal.\/} {\bf 14}~(4),
  736--766.

\bibitem[Umbanhowar {\em et~al.\/}(1996)Umbanhowar, Melo \&
  Swinney]{umbanhowar1996}
{\sc Umbanhowar, Paul~B., Melo, Francisco \& Swinney, Harry~L.} 1996 Localised
  excitations in a vertically vibrated granular layer. {\em Nature\/} {\bf
  382}~(29), 793--796.

\bibitem[video(2014)]{video3patch}
{\sc video} 2014
  \url{http://www.staff.uni-bayreuth.de/~bt240035/video3patch.wmv}.

\bibitem[Vislovich(1990)]{vislovich1990}
{\sc Vislovich, A.~N.} 1990 Phenomenological equation of static magnetization
  of magnetic fluids. {\em Magnetohydrodynamics\/} {\bf 26}~(2), 178--183.

\bibitem[Woods \& Champneys(1999)]{woods1999}
{\sc Woods, P.~D. \& Champneys, A.~R.} 1999 Heteroclinic tangles and homoclinic
  snaking in the unfolding of a degenerate reversible {H}amiltonian-{H}opf
  bifurcation. {\em Physica~D\/} {\bf 129}~(3-4), 147--170.

\end{thebibliography}
\end{document}